\documentclass[11pt,a4paper]{article}
\usepackage{latexsym}
\usepackage{amssymb}
\usepackage{amsmath}
\usepackage{amsfonts}
\usepackage{mathrsfs}
\usepackage{cite}
\usepackage{axodraw}
\usepackage{graphics}
\usepackage{graphicx,array}
\usepackage{mathrsfs}

\catcode`@=11
\renewcommand{\section}{\@startsection{section}{1}{0pt}{\medskipamount}
    {\medskipamount}{\large\bf}} \numberwithin{equation}{section}
\catcode`@=12

\newcommand{\be}{\begin{equation}}
    \newcommand{\ee}{\end{equation}}

\def\tr{{\rm tr}}
\def\tr{{\rm tr}\,}

\def\cN{{\cal N}}

\def\bea{\begin{eqnarray}}
    \def\eea{\end{eqnarray}}
\def\nn{\nonumber}

\def\cN{{\cal N}}

\def\f{\frac}

\def\tr{{\rm tr}\,}

\def\nn{\nonumber}

\def\d{\delta}

\def\g{\gamma}

\def\ve{\varepsilon}

\def\sB{\stackrel{\frown}{\square}}

\def\eq{\eqref}
\def\pr{\partial}
\def\nb{\nabla}

\numberwithin{equation}{section}

\begin{document}
    \begin{titlepage}

        \begin{center}
            \vspace{1cm}

{\bf \Large On two-loop divergences of effective action\\
\vspace{0.2cm}

in $6D$, ${\cal N}=(1,1)$ SYM theory}

            \vspace{1.5cm}

            {\bf
                I.L. Buchbinder\footnote{joseph@tspu.edu.ru }$^{\,a,b,c}$,
                E.A. Ivanov\footnote{eivanov@theor.jinr.ru}$^{\,b,d}$,
                B.S. Merzlikin\footnote{merzlikin@tspu.edu.ru}$^{\,a,e}$,
                K.V. Stepanyantz\footnote{stepan@m9com.ru}$^{\,f,b}$
            }
            \vspace{0.4cm}

            {\it
                $^a$ Center of Theoretical Physics, Tomsk State Pedagogical University, 634061, Tomsk,  Russia \\ \vskip 0.15cm
                $^b$ Bogoliubov Laboratory of Theoretical Physics, JINR, 141980 Dubna, Moscow region, Russia\\ \vskip 0.1cm
                $^c$ National Research Tomsk State University, 634050, Tomsk, Russia \\ \vskip 0.1cm
                $^d$ Moscow Institute of Physics and Technology, 141700 Dolgoprudny, Moscow region, Russia \\ \vskip 0.1cm
                $^e$ Laboratory of Applied Mathematics and Theoretical Physics,
                Tomsk State University of Control Systems and Radioelectronics, 634050 Tomsk, Russia\\ \vskip 0.1cm
                $^f$ Department of Theoretical Physics, Moscow State University, 119991 Moscow, Russia
            }
        \end{center}

        \vspace{0.4cm}

\begin{abstract}
We study the off-shell structure of the two-loop effective action in
$6D, \cN=(1,1)$ supersymmetric gauge theories formulated in
$\cN=(1,0)$ harmonic superspace. The off-shell effective action
involving all fields of $6D, \cN=(1,1)$ supermultiplet is
constructed by the harmonic superfield background field method,
which ensures both manifest gauge covariance and manifest
$\cN=(1,0)$ supersymmetry. We analyze the off-shell divergences
dependent on both gauge and hypermultiplet superfields and
argue that the gauge invariance of the divergences is consistent
with the non-locality in harmonics. The two-loop contributions to
the effective action are given by harmonic supergraphs with the
background gauge and hypermultiplet superfields. The procedure is
developed to operate with the harmonic-dependent superpropagators in
the two-loop supergraphs within the superfield dimensional
regularization. We explicitly calculate the gauge and the
hypermultiplet-mixed divergences as the coefficients of
$\frac{1}{{\varepsilon}^2}$ and demonstrate that the corresponding
expressions are non-local in harmonics.

\end{abstract}

    \end{titlepage}

\section{Introduction}

Quantum properties of gauge theories in higher dimensions attract a
lot of attention for a long time (see, e.g.,
\cite{Bern2005,Bern2010,Bern2012,Fradkin:1982kf,Marcus:1983bd,
Marcus:1984ei,Kazakov:2002jd,Howe:1983jm,Howe:2002ui,Bossard:2009sy,Bossard:2009mn,Bossard:2015dva,SerIvan}).
Although these theories are not renormalizable by power-counting
since the degree of divergence increases with a number of loops, the
direct calculations sometimes demonstrate miraculous cancelations of
some divergences. The situation becomes even more interesting in
supersymmetric theories due to additional restrictions imposed by
supersymmetry. In particular, some arguments were adduced that the
maximally extended ${\cal N}=8$ supergravity is finite to all orders
\cite{Kallosh:2010kk,Kallosh:2011dp,Bern:2018jmv}. In this
connection, it would be very interesting to study the ultraviolet
divergences of supersymmetric gauge theories in the space-times of
the dimension $D>4$ and reveal possible mechanisms of cancelations
of the divergences. One can expect that supersymmetry and,
especially, the maximally extended supersymmetry, is capable to
improve the ultraviolet behavior in supersymmetric higher
dimensional gauge theories.

In this paper we concentrate on  $6D,$  $\cN=(1,1)$ supersymmetric
Yang--Mills (SYM) theory. This theory is in many aspects similar to
$4D, \cN=4$ SYM theory, and so one can expect some similarities in
the structure of divergences in both theories. However, they
essentially differ in the UV domain. In contrast to $\cN=4$ SYM
theory, which is finite to all loops
\cite{Sohnius:1981sn,Grisaru:1982zh,Mandelstam:1982cb,Brink:1982pd,Howe:1983sr},
its $6D$ counterpart is non-renormalizable by power-counting.
Nevertheless, an extended supersymmetry provides the on-shell
finiteness of the theory up to two loops
\cite{Kazakov:2002jd,Bossard:2009mn,Bossard:2009sy,Bork:2015zaa}.
Recently we have shown that $6D, \cN=(1,1)$ gauge theory is one-loop
finite even off shell \cite{Buchbinder:2016url,Buchbinder:2017ozh}.
The modern methods of computing scattering amplitudes
\cite{Bork:2015zaa} demonstrate that UV divergences in $6D,
\cN=(1,1)$ SYM theory should start from the three-loop level in
accordance with the equation $D = 4 + 6/L$ valid for $L\ge 2$
\cite{Bern:1997nh,Bern:1998ug,Bern:2012di} (see also
\cite{Howe:2002ui}), which relates the space-time dimension $D$ to
the number of loops $L$ at which divergences appear in the
corresponding maximally supersymmetric Yang--Mills
theory\footnote{For recent advances of the amplitude techniques see
\cite{Carrasco:2021otn} and references therein.}.

When investigating the quantum properties of supersymmetric
theories, it is highly desirable to make use of the formulations and
quantization procedure which preserve as many supersymmetries as
possible. For example, $4D$, ${\cal N}=1$ supersymmetry is manifest
if a theory is formulated in ${\cal N}=1$ superspace (see e.g.
\cite{Gates:1983nr,West:1990tg,Buchbinder:1998qv}). Similarly, $4D$,
${\cal N}=2$ supersymmetry can be kept manifest at all stages of
calculations within the harmonic superspace approach
\cite{Galperin:1984av,Galperin:1985ec,Galperin:2001uw,Buchbinder:2001wy}.
This formalism is very convenient for analyzing the structure of
quantum corrections. Say, the finiteness beyond the one-loop
approximation \cite{Grisaru:1982zh,Howe:1983sr} can be proved much
easier in the harmonic superspace
\cite{Buchbinder:1997ib,Buchbinder:2015eva}. The similar harmonic
$6D$ superspace \cite{Howe:1985ar,Zupnik:1986da} can be efficiently
used for dealing with quantum $6D$, ${\cal N}=(1,0)$ supersymmetric
theories. If this method is applied to (the maximally extended)
$6D$, ${\cal N}=(1,1)$ SYM theory, then ${\cal N}=(1,0)$
supersymmetry is preserved even at the quantum level, while ${\cal
N}=(0,1)$ supersymmetry is hidden and can in general be broken by
quantum corrections.

In our previous work \cite{Buchbinder:2021unt} we initiated the
study of the two-loop divergent contribution to the effective action
of $6D$, $\cN=(1,1)$ SYM theory in the harmonic superfield formalism
and calculated the divergent contributions proportional to
$\frac{1}{{\varepsilon}^2}$ in the gauge superfield sector of the
model. In the present paper we will analyze the general structure of
the two-loop divergent contributions to the effective action in both
gauge and hypermultiplet sectors.

The UV properties of various $6D,\, \cN=(1,0)$ and $\cN=(1,1)$
theories in $6D$ harmonic superspace approach were studied in
\cite{Buchbinder:2016gmc,Buchbinder:2016url,Buchbinder:2017ozh,Buchbinder:2017gbs,
Buchbinder:2017xjb,Buchbinder:2018bhs,Buchbinder:2018lbd,Buchbinder:2019gfb,Buchbinder:2020tnc,Buchbinder:2020ovf,Ivanov:2005qf}.
The ${\cal N}=(1,1)$ SYM theory in this formulation amounts to
${\cal N}=(1,0)$ Yang--Mills supermultiplet minimally coupled to the
hypermutiplet in the adjoint representation of the gauge group. Note
that, in general, $\cN=(1,0)$ theories are plagued by quantum
anomalies
\cite{Townsend:1983ana,Smilga:2006ax,Kuzenko:2015xiz,Kuzenko:2017xgh},
but $\cN=(1,1)$ SYM theory is not anomalous. It was found that
$6D,\,\cN=(1,1)$ theory is off-shell finite in the one-loop
approximation in the Feynman gauge
\cite{Buchbinder:2016url,Buchbinder:2017ozh}. However, the off-shell
divergences still reappear  in the non-minimal gauges
\cite{Buchbinder:2019gfb} (although they vanish on shell). The
two-loop divergences in the hypermultiplet two-point Green functions
were shown to vanish off shell \cite{Buchbinder:2017gbs}. However,
the complete two-loop calculation in the harmonic superspace
approach has not yet been done. In the present paper we will
calculate the leading divergences and demonstrate that this theory
in the two-loop approximation is not finite off shell even in the
Feynman gauge. However, we will demonstrate that the divergences
disappear on shell, as it should be.

It is worth noting that although the harmonic superspace provides
the manifest $\cN=(1,0)$ supersymmetry at the quantum level, in the
Feynman supergraphs it produces harmonic-dependent propagators and
integrations over harmonics, the number of which increases with the
number of loops. For explicit calculations of such integrals, one
needs to develop the special methods which would allow to evaluate
the higher-loop harmonic supergraphs, based upon various nontrivial
identities for the  harmonic distributions.

In the present paper we work out  the manifestly $\cN=(1,0)$
supersymmetric approach to the calculation of the two-loop
divergences of the effective action in the theory under
consideration. Using this approach we will evaluate the leading
divergences proportional to $\frac{1}{{\varepsilon}^2}$ in the mixed
gauge-hypermultiplet sector and show that the result exactly matches
the one derived earlier in  \cite{Buchbinder:2021unt}.

The paper is organized as follows. In Section 2 we recall the
formulation of the six-dimensional $\cN=(1,1)$ SYM theory in
$\cN=(1,0)$ harmonic superspace and discuss details of its
superfield quantization. Section 3 is devoted to the power-counting
analysis of possible two-loop divergent contributions to the
effective action. We show that the possible counterterms, being, of
course, local in space-time, can generically be non-local in
harmonics. In Section 4 we describe the general structure of
divergent two-loop harmonic supergraphs, as well as  the methods of
their evaluations. Section 5 is devoted to applications of the above
methods to the explicit calculation  of the two-loop divergences in
the mixed gauge-hypermultiplet sector. Summary collects the main
results and briefly outlines the future directions of the study.
Some technical details are passed to the Appendices.

\section{The harmonic superspace formulation of $6D, \cN=(1,1)$ supersymmetric gauge theories}
\label{Section_Harmonic_Superspace}

We start by a brief overview of the $\cN=(1,0)$ harmonic superspace
formulation of $6D, \cN=(1,1)$ SYM theory. In such a formulation,
the theory enjoys  a manifest off-shell $\cN=(1,0)$ supersymmetry
and an additional hidden on-shell $\cN=(1,0)$ supersymmetry.
Together they form the complete $6D, \cN=(1,1)$ supersymmetry of the
classical theory. After quantization, we arrive at a quantum field
theory with preserving the manifest $\cN=(1,0)$ supersymmetry at all
stages of calculations. We recall the basic concepts of the harmonic
superspace following notations and conventions of our previous
papers \cite{Buchbinder:2016url,Buchbinder:2017ozh}.

The $6D, \cN=(1,0)$ harmonic superspace is parametrized by the coordinates $(z, u) = (x^M, \theta^a_i,u^{\pm i})$, where $x^M$, $M= 0,..,5$, are $6D$ Minkowski space-time coordinates, $\theta^a_i$ with
$a=1,..,4\,$, $i=1,2\,$, stand for  Grassmann variables. The extra bosonic coordinates $u^{\pm}_i$, $u^{+i}u^-_i =1$ (called harmonics) parametrize the coset $SU(2)/U(1)$ of the automorphisms
group $Sp(1) \sim SU(2)$ \cite{Howe:1985ar,Zupnik:1986da}. The analytic coordinates $\zeta = (x_{\cal A}^M, \theta^{\pm a})$ are defined as
 \be
x^M_{\cal A} \equiv x^M + \frac i2
\theta^{+a}(\gamma^M)_{ab}\theta^{-b},\qquad  \theta^{\pm a}=
u^\pm_k\theta^{ak},
 \ee
where $(\gamma^M)_{ab}$ are the antisymmetric $6D$ Weyl $\gamma$-matrices, $(\gamma^M)_{ab} = - (\gamma^M)_{ba}\,, (\widetilde{\gamma}^M)^{ab} =
\frac12\varepsilon^{abcd}(\gamma^M)_{cd}\,,$ with  $\varepsilon^{abcd}$ being the totally antisymmetric Levi-Civita tensor.

By definition, the analytic superfields are annihilated by the operator $D_a^+ = u^+_i D^i_a$, where $D^i_a$ are the spinor covariant derivatives.
Also we will need the covariant derivative $D_a^- = u^-_i D^i_a$ and the harmonic derivatives
 \bea
D^{\pm\pm}= u^{\pm i} \frac{\partial}{\partial u^{\mp i}},\qquad
\qquad D^0 = u^{+i} \frac {\partial}{\partial u^{+i}} - u^{-i} \frac
{\partial}{ \partial u^{-i}}.
 \eea
The spinor and harmonic derivatives satisfy the algebra
 \bea
    \{D^+_a,D^-_b\}=i(\gamma^M)_{ab}\partial_M\,, \qquad [D^{++}, D^{--}] = D^0,
    \qquad [D^{\pm\pm},D^{\pm}_a]=0\,, \qquad
    [D^{\pm\pm},D^{\mp}_a]=D^\pm_a\,.
 \eea
The integration measures over the full harmonic and analytic superspaces are defined, respectively, as
 \be
d^{14}z \equiv d^6x_{\cal A}\,(D^-)^4(D^+)^4,\quad d\zeta^{(-4)}
\equiv d^6x_{\cal A}\,du\,(D^-)^4 ,
 \ee
where
 \be
(D^{\pm})^4 = -\frac{1}{24} \varepsilon^{abcd} D^\pm_a D^\pm_b
D^\pm_c D^\pm_d\,.
 \ee

In the harmonic superspace approach the gauge field is a component of the analytic gauge superfield $V^{++}$. Also it is necessary
to introduce a non-analytic harmonic connection $V^{--}$ as a solution of the harmonic zero-curvature condition
\cite{Galperin:2001uw}
 \bea
D^{++} V^{--} - D^{--}V^{++} + i[V^{++},V^{--}]=0\,.  \label{zeroc}
 \eea
Using these superfields, we construct the gauge covariant harmonic
derivatives $\nb^{\pm\pm}= D^{\pm\pm} + i V^{\pm\pm}$. The spinor gauge covariant derivatives can be written as \cite{Ivanov:2005qf}
 \bea
\nb^+_a = D^+_a, \qquad \nb^-_a = D^-_a + i {\cal A}^-_a, \qquad
\nb_{ab} = \pr_{ab} + i {\cal A}_{ab}\,, \label{deriv}
 \eea
where $\nb_{ab} = \f12(\g^M)_{ab} \nb_M$, $\nb_M=\partial_M-iA_M\,$ and the superfield connections are defined
as
 \be
{\cal A}^-_a= i D^+_a V^{--}\,,\quad  {\cal A}_{ab} = \f12  D^+_a D^+_b
V^{--}. \label{OtherConn}
 \ee
The covariant derivatives \eq{deriv} satisfy the algebra
\begin{equation}
\{\nb_a^+,\nb^-_b\}=2i\nb_{ab}\,,\qquad
[\nb_c^\pm,\nb_{ab}]=\frac{i}2\ve_{abcd}W^{\pm\, d},\qquad [\nb_M,
\nb_N] = i F_{MN}\,,\label{alg2}
\end{equation}
where $W^{a\,\pm}$ is the superfield strength of the gauge supermultiplet,
 \bea
W^{+a}= -\frac{i}{6}\varepsilon^{abcd}D^+_b D^+_c D^+_d V^{--}, \qquad
W^{-a} = \nb^{--}W^{+a}\,. \label{Wstr}
 \eea

The classical action of $6D,\,\cN=(1,1)$ SYM model
$S_0[q^+\,,V^{++}]$ in the harmonic superspace is written as
\cite{Zupnik:1986da}
\begin{equation}\label{S0}
S_0 = \frac{1}{\rm f^2}\sum\limits^{\infty}_{n=2} \frac{(-i)^{n}}{n}
\tr \int d^{14}z\, du_1\ldots du_n \frac{V^{++}(z,u_1 ) \ldots
V^{++}(z,u_n ) }{(u^+_1 u^+_2)\ldots (u^+_n u^+_1 )}  - \frac{1}{2
{\rm f^2}} \tr \int d\zeta^{(-4)}\,  q^{+A} \nb^{++} q^{+}_A .
\end{equation}
The $SU(2)$ Pauli-G\"ursey indices are raised and lowered by the
rule $q^{+}_A =\epsilon_{AB} q^{+B}$, $\epsilon_{12}=1$. The
analytic superfields $V^{++}$ and $q^{+ A}$ belong to the adjoint
representation of the gauge group, i.e,
\begin{equation}
V^{++}=(V^{++})^It^{I}\,, \quad q^{+}_A=(q^{+}_A)^I t^{I}\,, \quad
[t^I,t^J]=if^{IJK} t^K\,, \quad \tr (t^{I}t^{J}) = \frac12 \d^{IJ},
\end{equation}
where $t^{I}$ are generators of the gauge algebra with the structure
constants $f^{IJK}$, and the covariant harmonic derivative is
defined as
\begin{equation}
\nb^{++}q^{+}_A = D^{++}q^{+}_A + i [V^{++},q^{+}_A].
\end{equation}
The classical equations of motion for the theory (\ref{S0}) have the
form
\begin{equation}
F^{++} +\frac{i}{2} [ q^{+A}, q^+_A]=0\,, \qquad \nb^{++}q^+_A =0\,,
\label{eqm}
\end{equation}
where we have defined the Grassmann-analytic superfield
 \bea
F^{++} = (D^+)^4 V^{--}\,, \quad D^{++} F^{++}=0\,.
\label{identity0}
 \eea

The action (\ref{S0}) is invariant under the manifest $\cN=(1,0)$
supersymmetry and an additional hidden $\cN=(0,1)$ supersymmetry.
The hidden supersymmetry mixes the gauge and hypermultiplet
superfields \cite{Bossard:2015dva},
\begin{equation}\label{Hidden}
\delta_{(0,1)} V^{++} = \epsilon^{+ A}q^+_A\,, \quad \delta_{(0,1)}
q^{+}_A = -(D^+)^4 (\epsilon^-_A V^{--})\,, \quad \epsilon^{\pm}_A =
\epsilon_{aA}\theta^{\pm a}\,.
\end{equation}
As a result, the action (\ref{S0}) is invariant under $6D,\,
\cN=(1,1)$ supersymmetry \footnote{Though the second supersymmetry
(\ref{Hidden}) is an off-shell symmetry of the action (\ref{S0}),
its correct closure with itself and with the manifest ${\cal
N}=(1,0)$ supersymmetry  is achieved only on shell.}. It
is also invariant under the superfield gauge transformations
\begin{equation}\label{gtr}
\d V^{++} = - \nb^{++}\lambda\,, \qquad \d q^{+}_A =
i[\lambda,q^{+}_A]\,,
\end{equation}
with a real analytic superfield $\lambda$ in the adjoint
representation as the transformation parameter. The non-analytic
gauge connection $V^{--}$ is transformed as
\begin{equation}\label{V--Transf}
\d V^{--} = - \nb^{--}\lambda\,,
\end{equation}
The harmonic zero-curvature condition \eqref{zeroc} is invariant
under the gauge transformations \eqref{gtr} and \eqref{V--Transf}.
The gauge transformations of the gauge connections
\eqref{OtherConn}, of the covariant strengths \eqref{Wstr} and of
the analytic superfield \eqref{identity0} can be easily found. In
particular, $F^{++}$ transforms homogeneously, like $Q^{+A}$, which
ensures the covariance of the equations of motion \eqref{eqm}.

For quantizing gauge theories, it is convenient to use the
background field method, which allows constructing the manifestly
gauge invariant effective action. For $6D,\, \cN=(1,0)$  SYM theory
in the harmonic superspace formulation this method was worked out in
ref.
\cite{Buchbinder:2016gmc,Buchbinder:2016url,Buchbinder:2017ozh}. In
many aspects it is similar to that for $4D, \,\cN=2$ supersymmetric
gauge theories \cite{Buchbinder:1997ya,Buchbinder:2001wy} (see also
the review \cite{Buchbinder:2016wng}).

Following the background field method we split the superfields
$V^{++}$ and $q^+_A$ into the sums of the background superfields
$V^{++}$, $Q^+_A$ and the quantum ones $v^{++}$, $q^+_A$,
\begin{equation}
V^{++}\to V^{++} + {\rm f} v^{++}\,, \qquad q^{+}_A \to Q^{+}_A +
{\rm f} q^{+}_A.
\end{equation}
Next, we expand the effective action in a power series in quantum
superfields and obtain a theory of the superfields $v^{++}, q^{+}_A$
in the background of the classical superfields $V^{++},Q^+_A$, which
are treated as functional arguments of the effective action.
According to refs.
\cite{Buchbinder:2016gmc,Buchbinder:2016url,Buchbinder:2017ozh} the
expression for the effective action can be written as
 \begin{equation}\label{Effective_Action}
e^{i \Gamma[V^{++},Q^+]} = \mbox{Det}^{1/2}\sB \int {\cal
D}v^{++}\,{\cal D}q^+\, {\cal D} b\,{\cal D} c\,{\cal D}\varphi\,
\exp\Big(iS_{\rm total}-i\tfrac{\delta \Gamma[V^{++},Q^+]}{\delta
V^{++}}\, v^{++}-i\tfrac{\delta \Gamma[V^{++},Q^+]}{\delta Q^{+}}\,
q^{+} \Big).
 \end{equation}
Here $$S_{\rm total} = S_{0}[q^+,v^{++};Q^+,V^{++}] + S_{\rm
gf}[v^{++};V^{++}] + S_{\rm  FP}[b,c,v^{++};V^{++}] + S_{\rm
NK}[\varphi;V^{++}]\,,$$ where $$S_{0}[q^+,v^{++};Q^+,V^{++}] =
S_{0}[Q^{+}+fq^{+}, V^{++}+fv^{++}].$$ The action $S_{\rm total}$
also includes the gauge-fixing term corresponding to the Feynman
gauge,
\begin{equation}\label{SGF}
S_{\rm gf}[v^{++}, V^{++}] = -\frac{1}{2}\tr \int d^{14}z\, du_1
du_2\,\frac{v_\tau^{++}(1)v_\tau^{++}(2)}{(u^+_1u^+_2)^2} +
\frac{1}{4}\tr \int d^{14}z\, du\, v_\tau^{++} (D^{--})^2
v_\tau^{++}
\end{equation}
(the index $\tau$ means the ``$\tau$-representation'', see eq.
\eqref{tau} below), the action for the fermionic Faddeev-Popov
ghosts ${\bf b}$ and ${\bf c}$, as well as the action for the
bosonic real analytic Nielsen-Kallosh ghost $\varphi$,
\begin{eqnarray}\label{FP}
S_{FP} &=&-\tr\int d\zeta^{(-4)}\, \nb^{++} {\bf b}\,
(\nb^{++}{\bf c} +i[v^{++}, {\bf c}]),\\
\label{NK} S_{\mbox{\scriptsize NK}} &=& -\frac{1}{2}\tr \int
d\zeta^{(-4)}\, \varphi ({\nb}^{++})^2\varphi.
\end{eqnarray}
On the subset of analytic superfields the operator
$\sB=\frac{1}{2}(D^+)^4(\nb^{--})^2$ in the expression
\eqref{Effective_Action} is reduced to the covariant super
d'Alembertian
\begin{eqnarray}\label{Box_First_Part}
\sB = \eta^{MN} \nabla_M \nabla_N + i W^{+a} \nabla^{-}_a + i F^{++}
\nabla^{--} - \frac{i}{2}(\nabla^{--} F^{++}),
\end{eqnarray}
where $\eta_{MN} = diag\,(1,-1,-1,-1,-1)$ is $6D$ Minkowski metric.

To remove the mixed terms containing both $v^{++}$ and $q^{+}_A$ in
the quadratic part of the action,
 \bea
S_2+S_{\rm gf} &=& -\frac{1}{2}\tr\int d\zeta^{(-4)}\, v^{++}\sB
v^{++}
-\f12 \tr\int d\zeta^{(-4)}\,{q}^{+ A}\nb^{++} q^{+}_{A} \nn \\
&&- \f{i}{2}\tr \int d\zeta^{(-4)}\Big\{ Q^{+ A}[ v^{++}, q^{+}_{
A}] + {q}^{+ A}[v^{++}, Q^{+}_{A}]\Big\}\,, \label{S2}
 \eea
we shift the quantum hypermultiplet superfield as
\begin{equation}\label{replac}
(q^{+A}_{1})^I= (h^{+A}_{1})^I + \int d \zeta^{(-4)}_2\,
\big(G^{(1,1)}{}^A{}_B(1|2)\big)^{IJ} f^{JKL} (v^{++}_{2})^K
(Q^{+B}_{2})^L\,.
\end{equation}
The newly defined  hypermultiplet $h^{+}_{A}$ has the propagator
\begin{equation}
G^{(1,1)}{}^A{}_B (1|2) = \langle h^{+A}_{1} h^+_{2B}\rangle = 2
\delta^A{}_B G^{(1,1)}(1|2),
\end{equation}
where we omitted the gauge group indices. The Green function
$G^{(1,1)}{(1|2)}$ satisfies the equation $\nabla^{++}
G^{(1,1)}(1|2) = \delta_{{\cal A}}^{(3,1)}(1|2)\,$ and has the
following explicit form \cite{Galperin:2001uw}
\begin{equation}\label{Green1}
G^{(1,1)}(1|2)^{IJ} = \big(\sB_1{}^{-1}\big)^{KL} (D^+_1)^4
(D^+_2)^4 \bigg((e^{ib_1})^{KI} (e^{ib_2})^{LJ}
\f{\d^{14}(z_1-z_2)}{(u^+_1u^+_2)^3}\bigg),
\end{equation}
where the bridge $b$ is defined as $V^{\pm\pm} = -i e^{ib}D^{\pm\pm}
e^{-ib}$ or, in more detail, $-if^{IJK} (V^{\pm\pm})^{I} = -i
(e^{ib})^{JL} D^{\pm\pm} (e^{-ib})^{LK}$ \cite{Galperin:2001uw}.
Using the bridge, one can define the $\tau$-representation for a
superfield $X$ according to the prescription
 \be
 X_\tau = e^{-ib} X e^{ib}. \label{tau}
 \ee

After performing the shift \eq{replac}, the quadratic part of the
action $S_2$ \eq{S2} splits into few terms, each being bilinear in
quantum superfields,
 \bea
 S_2+S_{gf} &=& \frac{1}{2} \tr\int d\zeta_1^{(-4)}\,d\zeta_2^{(-4)}\,
 v_1^{++}\Big\{ \sB \d^{(3,1)}_A(1|2)
 - 2 Q^{+A}_1 G^{(1,1)}(1|2) Q^{+}_{2A}\Big\}v_2^{++}
 \nonumber
 \\&& - \frac{1}{2} \tr \int d\zeta^{(-4)}\,  h^{+A} \nb^{++} h^{+}_A\,.  \label{S22}
 \eea
Then the propagator of the gauge superfield in \eq{S22} is expressed
as
\begin{eqnarray}\label{Green2}
G_{IJ}^{(2,2)}(1|2) &=& -2 (\sB_1{}^{-1})^{IJ} \d^{(2,2)}_{\cal
A}(1|2)
\nn\\
&& + 4 (\sB_1{}^{-1})^{IN} f^{N K L_1} (Q^{+A}_1)^K (Q^+_{2A})^M
f^{M L_2 P} G^{(1,1)}{(1|2)}^{L_1L_2} (\sB_2{}^{-1})^{PJ} +
\dots,\qquad
\end{eqnarray}
and that of the Faddeev-Popov ghosts as
\begin{equation}\label{Green0}
G^{(0,0)}(1|2) =  i \langle b(\zeta_1,u_1) c(\zeta_2,u_2) \rangle =
-(u_1^-u_2^-)G^{(1,1)}(1|2)\,.
\end{equation}
The Nielsen--Kallosh ghosts matter only in the one-loop
approximation and do not appear in the two-loop supergraphs
considered in this paper.

For calculating the two-loop quantum corrections we will need
vertices which are cubic and quartic in quantum superfields. The
presence of the non-trivial background hypermultiplet leads to the
modification of the interaction vertices compared  to ref.
\cite{Buchbinder:2017gbs}. Indeed, the substitution \eqref{replac}
obviously modifies the interacting vertices. Due to this
substitution, in addition to the ordinary 'local' vertices produced
by the corresponding terms in the classical action \eq{S0} (with the
formal replacement $q^+\to h^+$),
\begin{eqnarray} \label{vert1}
S^{(3)}_{\rm hyper} &=& \f{\rm f}{4}\int d\zeta^{(-4)}\,
f^{IJK} ({h}^{+A})^I (v^{++})^J (h^{+}_{A})^K\,,   \\
\label{vert2} S^{(3)}_{\rm SYM} &=& \f{i {\rm f} }{3}\tr\int d^{14}
z \prod_{a=1}^{3} du_a  \f{ v^{++}_1 v^{++}_2 v^{++}_3}{(u^+_1u^+_2)
(u^+_2u^+_3)(u^+_3u^+_1)}\,, \\
\label{vert3} S^{(4)}_{\rm SYM} &=& \f{{\rm f}^2}{4}\tr \int d^{14}
z \prod_{a=1}^{4} du_a \f{ v^{++}_1 v^{++}_2 v^{++}_3
v^{++}_4}{(u^+_1u^+_2) (u^+_2u^+_3)(u^+_3u^+_4)(u^+_4u^+_1)},
\end{eqnarray}
and those coming from the  Faddeev-Popov ghost action \eq{FP},
\begin{equation}\label{vert4}
S^{(3)}_{\rm ghost} = \f{f}{2}\int d\zeta^{(-4)}\, f^{IJK} (\nb^{++}
b)_I \, v^{++}_J c_K\,,
\end{equation}
we also encounter some new 'non-local' vertices involving the
hypermultiplet Green function $G^{(1,1)}(1|2)$,
\begin{eqnarray}\label{vert5}
\hspace*{-5mm} S^{(3)}_{\rm non-loc_1} &=&-{\rm f}\int
d\zeta_1^{(-4)} d\zeta_2^{(-4)} f^{IJK} G^{(1,1)}_{IL}(1|2) f^{LMN}
(Q^{+A}_2)^N \,(v^{++}_2)^M (v^{++}_1)^J (h^+_{1 A})^K\,, \\
\label{vert6} \hspace*{-5mm} S^{(3)}_{\rm non-loc_2} &=& -{\rm
f}\int d\zeta_1^{(-4)}d\zeta_2^{(-4)} d\zeta_3^{(-4)} f^{IJK}
G^{(1,1)}_{IL}(1|2) f^{LMN} \nn \\ && \times (Q^{+A}_2)^N
G^{(1,1)}_{KP}(1|3) f^{PQF} (Q^+_{3 A})^F \, (v^{++}_1)^J
(v^{++}_2)^M (v^{++}_3)^Q\,.
\end{eqnarray}
The presence of such types of vertices leads to additional
non-trivial diagrams, which we will consider in detail in the next
Sections.

\section{Power counting}

The superficial degree of divergence $\omega(G)$ for L-loop
supergraph G can be found by counting the degrees of momenta in the
loop integrals taking into account dimensions of the propagators and
the background hypermultiplet lines\footnote{The number of external
gauge lines has no impact on the index $\omega(G)$ because the
superfield $V^{++}$ is dimensionless.}. The result
\cite{Buchbinder:2016gmc} is
\begin{equation}
\label{sdd} \omega(G) =  2L - N_{Q},
\end{equation}
where $N_Q$ is a number of external hypermultiplet legs. Below we
investigate possible divergent contributions to the two-loop effective
action using the expression \eqref{sdd}.

\subsection{Possible two-loop divergences}
We begin with the general analysis of possible two-loop
contributions to the divergent part of the effective action defined
within the background field method. The power counting performed in
\cite{Buchbinder:2017ozh} leads to the following superficial degree
of divergence for the two-loop diagrams in the theory under
consideration
 \bea
 \omega_{\rm 2-loop}=4 - N_Q\,,
 \label{index}
 \eea
where $N_Q$ stands for a number of the hypermultiplet legs in the
diagrams and it should be even as the form of the classical action
indicates.\footnote{In the analysis below we take  all superfields
in the $\tau$-representation \cite{Galperin:2001uw}.}. In the
two-loop approximation, $N_Q$ can be equal to $0,\,2,\,4.$ This
allows us to list all possible two--loop divergent contributions to
the effective action.

First of all we note that in the framework of the $\cN=(1,0)$
supersymmetric background field method the (dimensionally
regularized) divergences  of the effective action are automatically
gauge invariant and $\cN=(1,0)$ supersymmetic by construction.
Taking into account the space-time locality of the counterterms
together with $\cN=(1,0)$ supersymmetry, we conclude that the
divergences should be local in the space-time and Grassmann
coordinates. However, since the harmonics are dimensionless, the
power counting cannot guarantee a locality in harmonics. This means
that we could expect the supersymmetric and gauge invariant
divergences which contain the harmonic non-localities. We believe
that the existence of gauge-invariant functionals that are nonlocal
in harmonics is a general property of extended supersymmetric
theories formulated in harmonic superspace. An example of such a
functional is the initial classical action \eqref{S0}. Further we
shall consider the consequences of the relation (\ref{index}).\\

\noindent $\underline{N_Q = 0,\, \, \omega = 4}$. The corresponding divergent contribution to
the two-point Green function of the gauge superfield is of the fourth
order in momenta and is given by the full $\cN = (1,0)$ superspace
integral
 \bea
 \Gamma_1^{(2)} \sim \tr \int d^{14} z du\, V^{--}_{\mbox{\scriptsize linear}} \square^2 V^{++}.
 \label{dim1}
 \eea
Passing to the analytic subspace in \eq{dim1} and restoring
terms with higher powers of gauge superfield on the ground of gauge
invariance, we obtain the expression
 \bea
 \Gamma_1^{(2)} \sim \tr\int d^{14}z \f{du_1du_2}{(u_1^+u_2^+)^2}
 F^{++}_{1} F^{++}_{2}\,,
 \label{dim3_1}
 \eea
where $F^{++}_{1, \tau}=F^{++}_{\tau}(z,u_1)$ and $[F^{++}]=[m^2]$.
In Appendix A we show that the expression \eq{dim3_1} can be
identically rewritten in the simpler form which is local in
harmonics,
 \bea
 \Gamma_1^{(2)} \sim \tr \int d \zeta^{(-4)}\,F^{++} \sB F^{++}\,.
\label{dim2}
 \eea
The expression \eqref{dim2} is local in the analytic coordinate $\zeta$
and harmonic variables $u$ in the analytic superspace. Just such an
expression was used in our previous paper \cite{Buchbinder:2021unt}.\\

\noindent $\underline{N_Q = 2\,, \, \omega = 2}$. The possible divergent contribution to
the effective action in the lowest order in $V^{++}$ includes only
one d'Alembertian operator, when written in the full $\cN=(1,0)$
superspace
 \bea
 \Gamma_2^{(2)} \sim \tr \int d^{14} z du\,
 V^{--}_{\mbox{\scriptsize linear}} \square [Q^{+A},Q^+_A]\,.
 \label{Gam2_1}
 \eea
There exist three following gauge invariant expressions reduced to
(\ref{Gam2_1}) in the lowest order in $V^{++}$ \bea
 {\cal G}_{2}^{(2)} &\sim&  \tr \int d^{14} z \f{du_1 du_2}{(u_1^+ u_2^+)^2}
 F^{++}_{1}[Q_{2}^{+A},Q^+_{2 A}] \,, \label{dim5} \\
 {\cal G}_{3}^{(2)} &\sim& \tr \int d^{14} z \f{du_1 du_2 du_3}{(u_1^+ u_2^+) (u_1^+u_3^+)}
 F^{++}_{1}[Q_{2}^{+A},Q^+_{3 A}] \label{dim6}\,, \\
 {\cal G}_{4}^{(2)} &\sim& \tr
 \int d^{14}z  \f{d u_1 du_2 du_3\,(u_1^-u_2^+)}{(u_1^+u_2^+)(u_2^+u_3^+)} F^{++}_{1,\tau} [Q^{+A}_{2,\tau}, Q^+_{3A,\tau}]\,.
 \label{dim7}
 \eea
Note that the hypermultiplet commutators appear because the
hypermultiplet is taken in the adjoint representation and they are
necessary for ensuring gauge invariance. In the Appendix A we show
that the expression \eqref{dim5} can be identically rewritten in the
simpler
 form local in harmonics
 \bea
 \Gamma_2^{(2)} \sim \tr \int d \zeta^{(4)} F^{++} \sB [Q^{+A},Q^+_A]\,,
 \label{Gam2}
 \eea
The expression (\ref{dim6}) and (\ref{dim7}) are gauge-invariant
functionals which are non-local in harmonics.\\

\noindent $\underline{N_Q = 4,\, \omega = 0}$. The possible lowest degree divergent
contribution to the four-point Green function of the hypermultiplet
superfield in the full $\cN=(1,0)$ superspace has the form
 \bea
 \Gamma_3^{(2)} \sim \tr \int d^{14} z du\, [Q^{+A},Q^+_A] (D^{--})^2 [Q^{+B},Q^+_B]\,.
 \label{4}
 \eea
The possible gauge invariant expression in the analytic subspace
corresponding to (\ref{4}) can be written as
 \bea
 \Gamma_3^{(2)} \sim \tr \int d \zeta^{(-4)}\, [Q^{+A},Q^+_A] \sB
 [Q^{+B},Q^+_B]\,.
 \label{Gam3}
 \eea
However, like in the previous case, one can construct more gauge
invariant expressions of the same mass dimensions as \eqref{Gam3},
such that they  are non-local in harmonics (see, e.g., a  recent
discussion in \cite{BBIMS22}). As an example, we quote the operator
 \bea
 {\cal G}_{5}^{(2)} &\sim& \tr \int d^{14} z \prod_{a=1}^4 du_a\,
 \f{[Q^{+A}_{1,\,\tau},Q^{+}_{2A,\,\tau}][Q^{+A}_{3,\,\tau},Q^{+}_{4A,\,\tau}]}{(u_2^+ u_3^+)(u_4^+ u_1^+)}\,,
 \label{dim10}
 \eea
which in the lowest order in $V^{++}$ coincides with \eqref{Gam3}.

\subsection{On-shell structure}
Here we consider the structure of the divergences for the
background superfields satisfying the classical equations of motion
(\ref{eqm}). The classical theory \eq{S0} possesses the hidden
on-shell $\cN=(0,1)$ supersymmetry \eqref{Hidden}, while the quantum
effective action does not preserve such a supersymmetry since the
gauge fixing conditions used for the construction of the effective
action preserve only the manifest $\cN=(1,0)$ supersymmetry.
However, for  the background superfields satisfying the classical
equations of motion, the hidden supersymmetry is restored.
Therefore, studying the on-shell divergences amounts  in fact to
the  study of restrictions on the divergences imposed by hidden
$\cN=(0,1)$ supersymmetry.

In general, the off-shell two-loop divergent contributions to the
effective action may include all possible terms of the form ${\cal
G}^{(2)}_i$, $i=1,..,5$. Let us examine the corresponding on-shell
contributions. First we have to note that all the contributions
${\cal G}^{(2)}_i$ should exactly yield one of the expression
$\Gamma^{(2)}_i$, $i=1,2,3$, if the equations of motion for the
hypermultiplet are satisfied (see Appendix A for details). Hence, we
can write
 \bea
 &&\Gamma^{(2)}_{\rm div} = \tr \int d \zeta^{(-4)}\Big(
 c_1 F^{++} \sB F^{++} + c_2 i \, F^{++} \sB [Q^{+A},Q^+_A] + c_3 \, [Q^{+A},Q^+_A] \sB [Q^{+B},Q^+_B] \Big)\nn \\
 && + \, \text{terms vanishing on e.o.m. for $Q^+$}\,,
 \label{Gdiv1}
 \eea
which is in the full agreement with our previous analysis in
\cite{Buchbinder:2021unt}. This expression involves three arbitrary
coefficients, each being proportional to $\frac{1}{{\varepsilon}^2}$
and to the dimensionful  coupling constant $f^2$ with dimensionless
numerical coefficients. Fixing the coefficients requires the
explicit calculations of the quantum corrections.

The hidden $\cN=(0,1)$ supersymmetry \eq{eqm} additionally restricts
the structure of the divergent contribution \eq{Gdiv1} and leads to
the absence of two-loop divergences on shell (see e.g.
\cite{Bossard:2015dva,SerIvan},\cite{Bork:2015zaa} and the
references therein). As we pointed out, the hidden $\cN=(0,1)$
supersymmetry means that the background superfields solve the
classical equations of motion \eqref{eqm}. In this case there arises
a connection between the coefficients $c_i$, $i=1,2,3,$ in
\eq{Gdiv1}.

Omitting the terms proportional to the hypermultiplet equation of
motion, we can  rewrite the action \eq{Gdiv1} as
 \bea
 \Gamma^{(2)}_{\rm div} &=& \tr \int d \zeta^{(-4)}
 \Big( c_1 E^{++} \sB E^{++} +(c_2-c_1)i F^{++} \sB [Q^{+A},Q^+_A] \nn \\
 &&\qquad \qquad +(c_3+\tfrac14 c_1) [Q^{+A},Q^+_A] \sB [Q^{+B},Q^+_B]
 \Big),
 \label{Gdiv2}
 \eea
where $E^{++} = F^{++}+\frac{i}{2}[Q^{+A}, Q^+_A]$ is the left hand
side of the gauge multiplet equation of motion \eq{eqm}. The
requirement that the expression \eq{Gdiv2} vanishes on-shell,
$E^{++}=0,$  leads to the relation between the coefficients
 \bea\label{c3equation}
2 c_2 + 4 c_3 = c_1\,. \label{con}
 \eea
The  leading contribution to the coefficient $c_1$ was calculated in
our previous work \cite{Buchbinder:2021unt}
 \bea
 c_1=\frac{{\rm f^2} (C_2)^2}{8(2\pi)^6\varepsilon^2}\,, \qquad \varepsilon \to 0\,,
 \label{coefic}
 \eea
for the case $Q^+_A=0$ that corresponds to the divergent
contribution in the pure gauge superfield sector. One of the
coefficients $c_2,\,c_3$  in (\ref{Gdiv1}) remains arbitrary. Hence,
to check the relation (\ref{Gdiv1}) which was obtained solely on the
ground of the power-counting arguments, it suffices to calculate
only one of these coefficients, e.g., $c_2$. The last one will be
fixed by eq. \eq{con}. In what follows, we consider those
contributions which include only the term $F^{++} Q^{+A} Q^+_A$ in
the divergent supergraphs and explicitly calculate the coefficient
$c_2$.

\section{Analysis of the two-loop harmonic supergraphs}

In this section we will discuss evaluation of the leading two-loop
divergences in the theory under consideration.

The supergraphs contributing to the two-loop effective action are
presented in Figs. \ref{Fig1}, \ref{Fig2}, and \ref{Fig3}. The first
two contain the conventional two-loop diagrams of six-dimensional
$\cN=(1,1)$ SYM theory \cite{Buchbinder:2021unt}. Fig. \ref{Fig3}
includes a new type of diagrams which are induced by the non-local
shift \eqref{replac}. In the course of calculations we do not assume
any restriction on the background gauge multiplet and hypermultiplet
and perform the analysis in a manifestly gauge invariant form.

\begin{figure}[tb]
\begin{center}
\begin{picture}(220,60)(0,0)
\PhotonArc(15,40)(30,0,360){2}{18}
\PhotonArc(75,40)(30,0,360){2}{18} \Vertex(45,40){2}
\Text(45,1)[]{$\Gamma_{\rm I}$}
\PhotonArc(190,40)(30,0,180){2}{9} \Photon(220,40)(160,40){2}{7}
\PhotonArc(190,40)(30,180,360){2}{9} \Vertex(160,40){2}
\Vertex(220,40){2}\Text(190,1)[]{$\Gamma_{\rm II}$}
\end{picture}
\caption{Two-loop Feynman supergraphs with the gauge
self-interactions vertices.\label{Fig1}}
\end{center}
\end{figure}
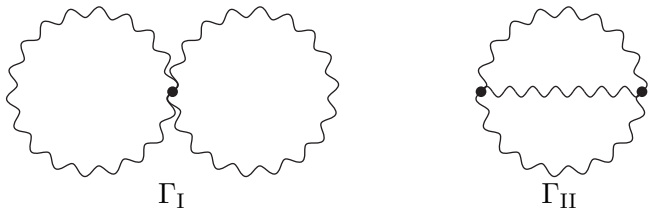
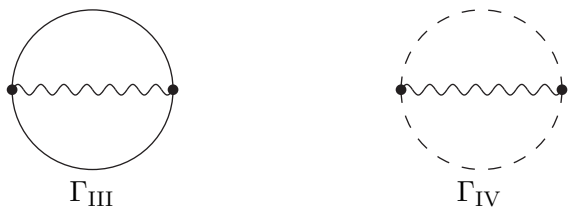
\begin{figure}[tb]
        \begin{center}
            \begin{picture}(220,60)(0,0)
                \CArc(45,40)(30,0,180) \Photon(15,40)(75,40){2}{7}
                \CArc(45,40)(30,180,360) \Vertex(75,40){2} \Vertex(15,40){2}
                \Text(45,1)[]{$\Gamma_{\rm III}$}
                \DashCArc(190,40)(30,0,180){5} \Photon(220,40)(160,40){2}{7}
                \DashCArc(190,40)(30,180,360){5} \Vertex(160,40){2}
                \Vertex(220,40){2}\Text(190,1)[]{$\Gamma_{\rm IV}$}
            \end{picture}
            \caption{Two-loop Feynman supergraphs with the hypermultiplet and ghosts
                vertices.\label{Fig2}}
        \end{center}
    \end{figure}

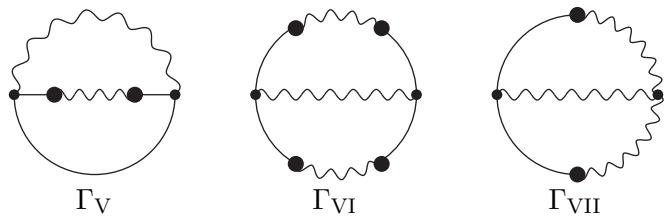
\begin{figure}[tb]
\begin{center}
\begin{picture}(240,70)(0,0)
\PhotonArc(30,40)(30,0,180){2}{8}  \Vertex(15,40){3}
\Vertex(45,40){3}  \Photon(15,40)(45,40){2}{4} \Line(0,40)(15,40)
\Line(45,40)(60,40) \CArc(30,40)(30,180,360) \Vertex(60,40){2}
\Vertex(0,40){2} \Text(30,0)[]{$\Gamma_{\rm V}$}
\CArc(120,40)(30,120,240) \Photon(90,40)(150,40){2}{7}
\CArc(120,40)(30,-60,60)\Vertex(90,40){2}
\PhotonArc(120,40)(30,60,120){2}{4}
\PhotonArc(120,40)(30,240,300){2}{4} \Vertex(105,65){3}
\Vertex(137,65){3} \Vertex(105,14){3} \Vertex(137,14){3}
\Vertex(150,40){2}\Text(120,0)[]{$\Gamma_{\rm VI}$}
\CArc(210,40)(30,90,270) \Photon(240,40)(180,40){2}{7}
\Vertex(180,40){2} \PhotonArc(210,40)(30,-90,90){2}{13}
\Vertex(210,70){3} \Vertex(210,10){3}
\Vertex(240,40){2}\Text(210,0)[]{$\Gamma_{\rm VII}$}
\end{picture}
\caption{Two-loop Feynman supergraphs with the new 'non-local'
vertices.\label{Fig3}}
\end{center}
\end{figure}

The graph of '$\infty$' topology (the contribution $\Gamma_{\rm I}$)
contains a vertex corresponding to the quartic interaction. In the
theory under consideration it is given by eq. \eq{vert3}. The graphs
of the '$\Theta$' topology are generated by cubic interactions. In
the considered $\cN=(1,1)$ theory they are given by eqs. \eq{vert1},
\eq{vert2}, \eq{vert4}, \eq{vert5}, and \eq{vert6}. The
corresponding contributions are denoted by $\Gamma_{\rm II} -
\Gamma_{\rm VII}$. Two cubic vertices \eqref{vert5} and
\eqref{vert6} appear due to the non-trivial hypermultiplet
background $Q^+$ and the non-local shift \eqref{replac}. These
vertices involve the background hypermultiplet $Q^+$ and the Green
function $G^{(1,1)}$.

In the supergraphs presented in Figs. \ref{Fig1} and \ref{Fig2} the
gauge propagators $G^{(2,2)}$, the hypermultiplet propagators
$G^{(1,1)}$, and the Faddeev--Popov ghost propagators $G^{(0,0)}$
are depicted by wavy, solid, and dashed lines, respectively. The
bold dotes in Fig. \ref{Fig3} denote the vertices which include the
background hypermultiplet $Q^+$.

It is known that the theory under consideration is finite at one
loop. Therefore, there is no need to renormalize the one-loop
subgraphs in the two-loop superdiagrams.

\paragraph{$\Gamma_{\rm I}$ diagram.}
The analytic expression corresponding to the diagram of the
`$\infty$' topology $\Gamma_{\rm I}$ presented in Fig. \ref{Fig1} is
written as
\begin{eqnarray}\label{eighth-1}
&& \Gamma_{\rm I} = - 2 {\rm f}^2\,{\rm tr} (t^I t^J t^K t^L)
\int {\rm d}^{14}z \int \prod_{a=1}^{4} du_a \nn \\
&& \times \left\{ \frac{G^{(2,2)}_{IJ} (z,u_1 ; z, u_2) \,
G^{(2,2)}_{KL} (z, u_3; z, u_4) } {(u^+_1u^+_2)(u^+_2u^+_3)(u^+_3
u^+_4) (u^+_4 u^+_1)} +\f12 \frac{ G^{(2,2)}_{IJ} (z,u_1 ; z, u_3)
\, G^{(2,2)}_{KL} (z, u_2; z, u_4) } {(u^+_1u^+_2)(u^+_2u^+_3)(u^+_3
u^+_4) (u^+_4 u^+_1)} \right\}.
\end{eqnarray}
The expression \eq{eighth-1} contains two gauge multiplet Green
functions $G^{(2,2)}$ in the limit of coincident $z$ points. Each
expression for $G^{(2,2)}$ contains a harmonic $\delta$-function,
but the first term in \eq{eighth-1} contains the expressions $(u_1^+
u_2^+)$ and $(u_3^+ u_4^+)$ in the denominator which produce a
singularity. As was demonstrated in \cite{Buchbinder:2021unt} it is
this term that contains the divergent  contribution of the diagram
$\Gamma_{\rm I}$. We discuss it in detail in the next Section.

The second term in \eq{eighth-1} also includes the gauge multiplet
Green function in the limit of coincident points, but the
denominator in this case does not vanish. Because of this, the
second term in eq. (\ref{eighth-1}) does not contribute to the
divergent part of the effective action. To demonstrate this we
consider the case $Q^+_A=0$ and  use the ``long'' form
\cite{Galperin:2001uw} of the Green function $G^{(2,2)}$
\eqref{Green2}:
\begin{eqnarray}\label{Gg}
G_{IJ}^{(2,2)}(1|2) &=& -\big({\sB_1{}^{-2}}\big)^{KL}
(D^+_1)^4(D^+_2)^4
\Big((e^{ib_1})^{IK} (e^{ib_2})^{JL} \d^{14}(z_1-z_2)\, (D^{--}_1)^2\d^{(2,-2)}(u_1,u_2)\Big)\, \label{id1}\\
&=& -\big({\sB_1{}^{-2}}\big)^{KM} (D^+_1)^4 \bigg[\Big\{
(D^{-}_1)^4 (u_1^+u_2^+)^4
-\Omega^{--}_1 (u_1^-u_2^+)(u_1^+u_2^+)^3  + \sB_1 (u_1^-u_2^+)^2(u_1^+u_2^+)^2 \nn\\
&& -i F_1^{++}(u_1^-u_2^+)^3(u_1^+u_2^+)\Big\}^{ML}\,
\Big((e^{ib_1})^{KI} (e^{ib_2})^{LJ} \d^{14}(z_1-z_2)\,
(D^{--}_1)^2\d^{(2,-2)}(u_1,u_2)\Big)\bigg]\,,\nn
\end{eqnarray}
where $ \Omega^{--} \equiv i\nb^{ab}\nb^-_a \nb^-_b - W^{-a}\nb^-_a
+ \frac{1}{4} (\nb^-_a W^{- a})~.$ Then, due to the identity
\begin{eqnarray}
(u_1^+u_2^+)^q (D^{--}_1)^k \delta^{(2,-2)}(u_1,u_2) = 0, \quad  q >
k\,,
\end{eqnarray}
the non-zero contribution emerges only for $q=k$, in particular,
 \bea
  (u_1^+u_2^+)^2 (D^{--}_1)^2 \delta^{(2,-2)}(u_1,u_2)&=&
2\, \delta^{(0,0)}(u_1,u_2)\,. \label{id11}
 \eea
Hence, we need to collect additional $D^{--}$ operators from the
d'Alembertian operator in the denominator in the first and second
terms of the expression \eq{Gg}. For this purpose, we consider the
Green function \eq{Gg} in the coincident point $z_2\to z_1$ limit
and collect the fourth power of $D^-_a$ from the inverse $\sB$
operator to annihilate the Grassmann $\delta$-function
$\delta^{8}(\theta_1-\theta_2)$,
\begin{eqnarray}
\f{1}{\sB}=\f{1}{\square + W^{+a} D^-_a + \dots} \sim
\f{(W^{+})^4(D^-)^4}{\square^5}.
\end{eqnarray}
This implies that the second term in \eqref{eighth-1} gives only
finite contributions, and so the divergent contributions can come
only from the first term in eq. \eqref{eighth-1}.

\paragraph{$\Gamma_{\rm II}$ diagram.}The analytic expression
for the two-loop diagram $\Gamma_{\rm II}$ (of the `$\Theta$'
topology) presented in Fig. \ref{Fig1} is constructed using the
cubic gauge superfield vertex \eq{vert2}:
\begin{eqnarray}
\Gamma_{\rm II} &=& -\f{{\rm f}^2}{6} \int d^{14}z_1
\,d^{14}z_2\prod_{a=1}^{6}
du_a \,f^{I_1J_1K_1}  f^{I_2J_2K_2} \nn \\
&&\times \frac{G^{(2,2)}_{I_1 I_2} (z_1,u_1 ; z_2, u_4) \,
G^{(2,2)}_{J_1 J_2} (z_1, u_2; z_2, u_5) \, G^{(2,2)}_{K_1 K_2}
(z_1,u_3 ; z_2, u_6)} {(u^+_1u^+_2)(u^+_2u^+_3)(u^+_3 u^+_4)
\,(u^+_4u^+_5)(u^+_5u^+_6)(u^+_6 u^+_1)}.
\end{eqnarray}
We substitute the explicit expressions for the Green function
$G^{(2,2)}$ and integrate by parts with respect to one of the
$(D^+)^4$ factors. After this it is necessary to calculate the
harmonic integrals over $u_4,u_5,u_6$ using the corresponding
delta-functions which come from the propagators. As a result, we
obtain
\begin{eqnarray}\label{Diagram2}
\Gamma_{\rm II} &=& -\f{{\rm f}^2}{6} \int d^{14}z_1
\,d^{14}z_2\prod_{a=1}^{3} du_a \, \f{f^{I_1J_1K_1}
f^{I_2J_2K_2}}{(u^+_1u^+_2)^2(u^+_2u^+_3)^2(u^+_3 u^+_1)^2}
\, (\sB{}^{-1})_{I_1I_2} \delta^{14}(z_1-z_2) \nn \\
&& \qquad\qquad\quad \times (D^{+}_1)^4 \Big( (\sB{}^{-1})_{J_1J_2}
(D_2^+)^4\delta^{14}(z_1-z_2) (\sB{}^{-1})_{K_1K_2}
(D_2^+)^4\delta^{14}(z_1-z_2) \Big).\quad
\end{eqnarray}
After integrating over $\theta_2$ using the Grassmann delta-function
we are left with the coincident $\theta_2 \to \theta_1$ limit in the
two remaining delta-functions. As in the previous case, we need to
annihilate these Grassmann delta-functions in the coincident
$\theta$-point limit, acting on each of them by $(D^\pm)^8$-factors.
There are three explicit $(D^+)^4$ factors in the expression
(\ref{Diagram2}). Therefore, the remaining $(D^-)^4$ factor should
be taken from the inverse $\sB$ operators. However, doing so, we
will produce an extra operator $(\partial^2)^{-4}$. This implies
that the expression for the  superdiagram considered can give only
finite contribution to the effective action.

\paragraph{$\Gamma_{\rm III}$ and  $\Gamma_{\rm IV}$ diagrams.}
As in the previous consideration for the $Q^+=0$ case
\cite{Buchbinder:2021unt}, the superdiagrams $\Gamma_{\rm III}$ and
$\Gamma_{\rm IV}$ presented in Fig. \ref{Fig2} exactly cancel each
other due to ${\cal N}=(0,1)$ supersymmetry (requiring $Q^{+A}$ to
be in the adjoint representation) even for the non-trivial $Q^+$
background\footnote{See ref. \cite{Kuzenko:2004sv}, where a similar
situation was also observed in $4D$, $\cN=4$ SYM theory.}. Indeed,
the sum of these two superdiagrams is given by the expression
\begin{eqnarray}\label{Gamma1}
\Gamma_{\rm III}+\Gamma_{\rm IV} &=& {\rm f^2}\int d\zeta_1^{(-4)}
d\zeta_2^{(-4)}\,
\Big(1+(u^+_1 u^-_2)(u^-_1 u^+_2)\Big)   \nn \\
&& \times f^{I_1 J_1 K_1} f^{I_2 J_2 K_2} G^{(2,2)}_{I_1 I_2}(1|2)
\, G^{(1,1)}_{J_1 J_2}(1|2) \, G^{(1,1)}_{K_1 K_2}(1|2).
\end{eqnarray}
Applying the identity
\begin{equation}
1+(u^+_1 u^-_2)(u^-_1 u^+_2) = (u^+_1 u^+_2)(u^-_1 u^-_2)\,
\end{equation}
we cast the contribution \eq{Gamma1} in the form
\begin{eqnarray} \label{Gamma11}
\Gamma_{\rm III}+\Gamma_{\rm IV}  &=& {\rm f^2}
\int d\zeta_1^{(-4)} d\zeta_2^{(-4)}\, (u^+_1 u^+_2)(u^-_1 u^-_2)  \nn  \\
&& \qquad\qquad\times f^{I_1 J_1 K_1} f^{I_2 J_2 K_2} G^{(2,2)}_{I_1
I_2}(1|2) \, G^{(1,1)}_{J_1 J_2}(1|2) \, G^{(1,1)}_{K_1 K_2}(1|2)
=0\,.
\end{eqnarray}

\paragraph{$\Gamma_{\rm V}$, $\Gamma_{\rm VI}$ and $\Gamma_{\rm VII}$ diagrams.}
Now, let us consider the additional superdiagrams $\Gamma_{\rm V}$,
$\Gamma_{\rm VI}$ and $\Gamma_{\rm VII}$ in Fig. \ref{Fig3}. These
diagrams are the modification of the divergent diagram $\Gamma_{\rm
III}$ in Fig. \ref{Fig2}, where the local cubic vertex \eqref{vert1}
is replaced by the non-local shifted vertices \eqref{vert5} and
\eqref{vert6}. The diagrams $\Gamma_{\rm III}$ and $\Gamma_{\rm IV}$
are divergent but cancel each other due to $\cN=(0,1)$
supersymmetry, see eq. \eqref{Gamma11}. Hence, the additional
diagrams in Fig.\ref{Fig3}, which appear after the non-local change
of variables \eqref{replac}, diverge. Since the substitution
\eqref{replac} involves only the quantum hypermultiplet and does not
act on the ghost superfields, the additional diagrams $\Gamma_{\rm
V}-\Gamma_{\rm VII}$ cannot be canceled.

We have to emphasize that we collected only divergent contributions
in the mixed gauge-hypermultiplet sector, such as $F^{++}Q^{+A}
Q^+_A$. The non-local vertices \eq{vert5} and \eq{vert6} include the
background hypermultiplet $Q^+_A$, which is depicted by the bold
dots in Fig.3 for diagrams $\Gamma_{\rm V}-\Gamma_{\rm VII}$.
However, the diagram $\Gamma_{\rm VI}$ contains at least four
background hypermultiplets and so is beyond our consideration. The
remaining diagrams $\Gamma_{\rm V}$ and $\Gamma_{\rm VII}$ include
the terms of interest, as implied by the power momentum counting and
contribute to the $F^{++}Q^{+A} Q^+_A$ part of the effective action.

\section{Calculation of the two-loop divergences}

As explained above, we are only interested in the divergent
contributions to the effective action of the form $F^{++}
Q^{+A}Q^+_A$, which appear solely from the superdiagrams
$\Gamma_{\rm I}$, $\Gamma_{\rm V}$ and $\Gamma_{\rm VII}$. In this
section we consider the calculation of the $1/{\varepsilon^2}$
divergences coming from these diagrams in some detail.

\paragraph{Divergent part of $\Gamma_{\rm I}$ diagram.} The analytic expression for the
superdiagram $\Gamma_{\rm I}$ presented in Fig.~\ref{Fig1} is
constructed similarly to the case $Q^+=0$ considered in
\cite{Buchbinder:2021unt}. The new peculiarity is that the vector
supermultiplet propagator \eqref{Green1} now includes a non-trivial
$Q^+$ dependence,
\begin{equation}\label{eighth}
\Gamma_{\rm I,\, div} = - 2 {\rm f}^2\,{\rm tr} (t^I t^J t^K t^L)
\int {\rm d}^{14}z \int \prod_{a=1}^{4} du_a \frac{G^{(2,2)}_{IJ}
(z,u_1 ; z, u_2) \, G^{(2,2)}_{KL} (z, u_3; z, u_4) }
{(u^+_1u^+_2)(u^+_2u^+_3)(u^+_3 u^+_4) (u^+_4 u^+_1)}.
\end{equation}
To calculate the divergent contribution to the effective action, we
substitute the expression \eqref{Green2} for the vector multiplet
Green function into the numerator in \eq{eighth}. Then, using the
identity
\begin{equation}\label{id2}
\bigg[ \f{1}{\sB_1^2}, \f{1}{(u_1^+u_2^+)}\bigg] =
\f1{\sB_1}\Big(\f{(u_1^-u_2^+)}{(u_1^+u_2^+)^2}F^{++}\Big)\f1{\sB_1^2}
+\f1{\sB_1^2}\Big(\f{(u_1^-u_2^+)}{(u_1^+u_2^+)^2}F^{++}\Big)\f1{\sB_1}\,
\end{equation}
and omitting terms that give finite contributions the expression
\begin{eqnarray}\label{Gg2}
\f{1}{(u_1^+u_2^+)}G_{IJ}^{(2,2)}(1|2) = \f{1}{(u_1^+u_2^+)} \Big( -
\f{2}{\sB_1} \d^{(2,2)}_{\cal A}(1|2) + \f{4}{\sB_1^2} Q^{+A}_1
G^{(1,1)}{(1|2)} Q^+_{2A} + \dots\Big)_{IJ}
\end{eqnarray}
can be calculated in the limit of coincident points. For this
purpose we note that in the leading order in $1/\varepsilon$, where
$\varepsilon\equiv 6-D \to 0$, in the expression considered one can
make the replacement
\begin{equation}\label{Int}
\f{1}{\square^3} \delta^6(x-x') \Big|_{x'=x} \to
\f{i}{(4\pi)^3\varepsilon}\,
\end{equation}
which is valid in the minimal subtraction scheme. Thus, we obtain
\begin{eqnarray}\label{Gg4}
\f{1}{(u_1^+u_2^+)}G^{(2,2)}(z_1,u_1|z_2,u_2)\Big|_{z_2\to z_1} &=&
\f{2i}{(4\pi)^3}\, \f{1}{\varepsilon}\, \Big( F^{++}_{1,\tau} \,
(u_1^-u_2^+)(u_1^+u_2^+)^2 (D^{--}_1)^2\d^{(2,-2)}(u_1,u_2) \nn \\
&& + 2\,Q^{+A}_{1,\tau} Q^+_{2A,\tau} \Big) + {\rm finite\
contributions} \,, \quad \varepsilon\to0\,,
\end{eqnarray}
where $\tau$-representation was introduced in Eq. \eqref{tau}.

To reproduce the result of \cite{Buchbinder:2021unt}, one can put
$Q^+=0$ in Eq. \eqref{Gg4} and substitute it to Eq. \eqref{eighth}.
Then we obtain
 \be
\Gamma^{F F}_{\rm I, div} = \f{{\rm f^2} (C_2)^2}{8(2\pi)^6
\varepsilon^2} \mbox{tr} \int d^{14}z\, du_1 du_2\, F_{1,\tau}^{++}
F_{2,\tau}^{++} \frac{1}{(u_1^+ u_2^+)^2}.
 \ee
Using the identity
    $$
F_{1,\tau}^{++} = \frac{1}{2} D_1^{++} D_1^{--} F_{1,\tau}^{++}\,,
    $$
and integrating by parts with respect to the derivative $D_1^{++}$, this expression can be rewritten as
 \bea
&& \mbox{tr} \int d^{14}z\, du_1 du_2\, F_{1,\tau}^{++}
F_{2,\tau}^{++} \frac{1}{(u_1^+ u_2^+)^2}
= - \frac{1}{2}\mbox{tr} \int d^{14}z\, du_1 du_2\, D_1^{--} F_{1,\tau}^{++} F_{2,\tau}^{++} D_1^{--} \delta^{2,-2}(u_1,u_2)\qquad \nonumber\\
&& =  \frac{1}{2}\mbox{tr} \int d^{14}z\, du\, (D^{--})^2
F_{\tau}^{++} F_{\tau}^{++} = \mbox{tr} \int d\zeta^{(-4)}\, du\,
F^{++} \sB F^{++}.
 \eea
Thus, the leading ($1/\varepsilon^2$) divergent part of the two-loop
diagram $\Gamma_{\rm I}$ depicted in Fig.1 depending only on the
background gauge superfield has the form
 \be
\label{Result1} \Gamma^{F F}_{\rm I, div} = \frac{{\rm f^2}
(C_2)^2}{8(2\pi)^6\varepsilon^2} \mbox{tr} \int d\zeta^{(-4)}\, du\,
F^{++} \sB F^{++},
 \ee
and determines the constant $c_1$ in Eq.\eqref{coefic}.

In the general $Q^+\neq 0$ case, we substitute the whole expression
\eq{Gg4} into eq. \eq{eighth} and obtain the leading
($1/\varepsilon^2$) divergent contribution coming from this diagram
in the form
\begin{eqnarray}\label{GammaI}
\Gamma^{\rm FQQ}_{\rm I, \,div} &=& -\f{i{\rm
f}^2(C_2)^2}{8(2\pi)^6\,\varepsilon^2}\tr \int d^{14}z \, \f{du_1
du_2 du_3}{(u^+_1 u^+_2)(u^+_3 u^+_1)} F^{++}_{1, \tau}
[Q^{+A}_{2,\tau}, Q^{+}_{3 A,\tau}]\,.
\end{eqnarray}
The expression \eqref{GammaI} generalizes the result obtained
earlier in \cite{Buchbinder:2021unt} for the case $Q^+=0$.

\paragraph{Divergent part of $\Gamma_{\rm V}$ diagram.}
The analytic expression for the diagram $\Gamma_{\rm V}$ reads
\begin{eqnarray}\label{GammaV-1}
\Gamma_{\rm V} &=& i {\rm f}^2 f^{I_1J_1K_1} f^{I_2J_2K_2}
f^{L_1M_1N_1} f^{L_2 M_2 N_2} \int \prod_{a=1}^{4} d\zeta_a^{(-4)}
\, (Q^{+A}_2)^{N_1} (Q^{+}_{4A})^{N_2}  \\
&& \times G^{(1,1)}_{I_1 L_1}(1|2)\, G^{(1,1)}_{I_2 L_2}(3|4)\,
G^{(2,2)}_{J_1 J_2}(1|3)\, G^{(2,2)}_{M_1 M_2}(2|4)\, G^{(1,1)}_{K_1
K_2}(1|3)\,. \nn
\end{eqnarray}

To single out  the divergent contribution in \eqref{GammaV-1}, we
should first substitute the explicit expressions for the Green
functions. In the case of the non-trivial hypermultiplet background,
the gauge superfield propagator $G^{(2,2)}$ is given by the
expression \eqref{Green2} involving additional contributions with
the hypermultiplet Green function. These terms contain one more
d'Alembert operator in the denominator and produce finite
contributions to the effective action. Thus, we should consider only
the first term in the expression \eqref{Green2}, which corresponds
to the standard propagator for the gauge superfield $v^{++}$. Then,
we integrate over $\zeta_3$ and $\zeta_4$ using the analytic
delta-functions $\delta^{(2,2)}_{\cal A}(1|3)$ and
$\delta^{(2,2)}_{\cal A}(2|4)$ coming from the gauge superfield
propagators. After making these steps,  we find the following result
for the leading divergences
\begin{eqnarray}
\Gamma^{\rm FQQ}_{\rm V,\, div} &=& 4i {\rm f}^2 f^{I_1J_1K_1}
f^{I_2J_2K_2} f^{L_1M_1N_1} f^{L_2 M_2 N_2} \int d\zeta_1^{(-4)}
d\zeta_2^{(-4)}
 (Q^{+A}_{2})^{N_1} (Q^{+}_{2A})^{N_2}
 \\
&& \times (\sB_1{}^{-1})^{J_1J_2} G^{(1,1)}_{K_1K_2}(1|2)\Big|_{2=1}
\, (\sB_2{}^{-1})^{M_1M_2} G^{(1,1)}_{I_1 L_1}{(1|2)}\,
G^{(1,1)}_{I_2 L_2}(1|2)\,. \nn
\end{eqnarray}
Next, we substitute the hypermultiplet Green function \eq{Green1},
restore the full superspace measure using $(D^{+})^4$ factors from
$\big(G^{(1,1)}_{(1|2)}\big)^{I_2 L_2}$, and integrate with the
Grassmann delta-function $\d^8(\theta_1 -\theta_2)$ coming from this
Green function. Taking the coincident points limit and using the
identity $(D_1^{+})^4 (D_2^{+})^4 \d^8(\theta_1
-\theta_2)\big|_{\theta_2=\theta_1} = (u_1^+u_2^+)^4$, we obtain
\begin{eqnarray}
\Gamma^{\rm FQQ}_{\rm V,\, div} &=& 8i{\rm f}^2 f^{I_1 J K_1} f^{I_2
J K_2} f^{I_1 M N_1} f^{I_2 M N_2}  \int d^{14} z_1 d^6 x_2
du_1du_2\,
\f{(Q^{+A}_{2,\tau})^{N_1} (Q^{+}_{2A,\tau})^{N_2}}{(u_1^+u_2^+)^2} \nn \\
&&\times f^{K_1K_3K_2} (F^{++}_{1,\tau})^{K_3} \f{1}{\square_1^3}
\delta^6(x_1-x_2)\big|_{x_2=x_1}\f{1}{\square_2\square_1}
\delta^6(x_1-x_2) \f{1}{\square_1} \delta^6(x_1-x_2)\,.
\end{eqnarray}
After calculating the (Euclidean) momentum integrals with the help
of the identity
\begin{equation}
\label{momint} \int \frac{d^Dk}{(2\pi)^D} \frac{d^Dl}{(2\pi)^D}
\frac{1}{k^6 l^4(k+l)^2} \to \frac{1}{(4\pi)^6}\cdot
\frac{1}{2\varepsilon^2} + O\Big(\frac{1}{\varepsilon}\Big)
\end{equation}
the result for the leading divergences can be written as
\begin{equation}\label{GammaV-2}
 \Gamma^{\rm FQQ}_{\rm V,\, div} = \f{2i{\rm f}^2 (C_2)^2}{(4\pi)^6 \varepsilon^2} \tr
 \int d^{14}z \f{d u_1 du_2}{(u_1^+u_2^+)^2} F^{++}_{1,\tau} [Q^{+A}_{2,\tau}, Q^+_{2A,\tau}]\,,
\end{equation}
where we have also used the relation $f^{ABC} f^{BKL} f^{CLN} = -
\f12 C_2 f^{AKN}$.

\paragraph{Divergent part of $\Gamma_{\rm V I I}$ diagram.}
The analytic expression for the diagram $\Gamma_{\rm V II} $
depicted on Fig.3 has the form
\begin{eqnarray}
\Gamma_{\rm VII} &=& - {\rm f}^2 f^{I_1J_1K_1} f^{I_2J_2K_2}
f^{L_1M_1N_1} f^{L_2 M_2 N_2} \int d^{14}z_1 \prod_{a=1}^{3} d u_a\,
\prod_{b=4}^{6} d\zeta_b^{(-4)} \f{(Q^{+A}_5)^{N_1}
(Q^{+}_{6A})^{N_2}}{(u^+_1u^+_2)
(u^+_2u^+_3)(u^+_3u^+_1)}  \nn \\
&& \times G^{(2,2)}_{I_1 J_2}(1|4)\, G^{(2,2)}_{J_1 M_1}(2|5)\,
G^{(2,2)}_{K_1 M_2}(3|6)\, G^{(1,1)}_{I_2 L_1}(4|5) G^{(1,1)}_{K_2
L_2}(4|6)\,. \label{GammaVII-1}
\end{eqnarray}
We are interested in its divergent part of the form
$F^{++}[Q^{+A},Q^+_A]$. The expression \eqref{GammaVII-1} already
contains the second power of the background hypermultiplet. Hence,
we take only the first term from the gauge superfield propagator
\eqref{Green2}. Then we take off the $(D^+)^4$ factors from each
Green function $G^{(2,2)}$ to restore the full superspace measures
over $z_4, z_5 $ and $z_6$. We also integrate over the Grassmann
variables $\theta_4,  \theta_5$ and $\theta_6$ and the corresponding
harmonic variables $u_4,u_5$ and $u_6$ using the $\delta$-functions
coming from the Green functions $G^{(2,2)}$. Then, taking the
hypermultiplet Green function $G^{(1,1)}$ in the limit of coincident
$\theta$'s, we obtain
 \begin{eqnarray}
\Gamma^{FQQ}_{\rm VII,\, div} &=& 8 {\rm f}^2 f^{I_1J_1K_1}
f^{I_1I_2K_2}
f^{L_1J_1N_1} f^{L_2 K_1 N_2} \int d^{14}z_1 d^{6}x_4 d^{6}x_5 d^{6}x_6 \f{d u_1 du_2 du_3}{(u^+_2u^+_3)} \,  \nn\\
&& \times (Q^{+A})^{N_1}(x_5,\theta_1,u_2) (Q^{+}_{A})^{N_2}(x_6,\theta_1,u_3) \nn \\
&&\times \f{1}{(u^+_1u^+_2)} \f{1}{\sB_{I_2 L_1}} (u^+_1u^+_2)
\delta^6 (x_4-x_5) \,
 \f{1}{(u^+_1u^+_3)} \f{1}{\sB_{K_2 L_2}} (u^+_1u^+_3) (x_4-x_6) \nn \\
&& \times \f{1}{\square_1 }\delta^6(x_1-x_4)
\f{1}{\square_1}\delta^6(x_1-x_5)\f{1}{\square_1}\delta^6(x_1-x_6)
\,. \label{GammaVII-2}
\end{eqnarray}

The operator $\sB$ (given by eq. \eqref{Box_First_Part}) does not
commute with harmonic factors due to the harmonic derivative
$\nabla^{--}$. In particular, using the identity
 \bea
 \Big[ \f{1}{\sB}, (u^+_1u^+_2)\Big] = -i (u^-_1u^+_2) \f{1}{\sB} F^{++} \f{1}{\sB}\,,
 \eea
we obtain for the divergent contribution
  \begin{eqnarray}
\Gamma^{FQQ}_{\rm VII,\, div} &=& 8 {\rm f}^2 f^{I_1J_1K_1}
f^{I_1I_2K_2}
f^{L_1J_1N_1} f^{L_2 K_1 N_2} \int d^{14}z_1 d^{6}x_4 d^{6}x_5 d^{6}x_6 \f{d u_1 du_2 du_3}{(u^+_2u^+_3)} \,  \nn\\
&& \times (Q^{+A})^{N_1}(x_5,\theta_1,u_2) (Q^{+}_{A})^{N_2}(x_6,\theta_1,u_3)\, (F^{++})^{N_3}(x_4,\theta_1,u_1) \nn \\
&&\times \bigg(\f{(u^-_1u^+_2)}{(u^+_1u^+_2)} \delta^{K_2 L_2}
f^{I_2 N_3 L_1}
\f{1}{\square^2_4} \delta^{6} (x_4-x_5) \f{1}{\square_4} \delta^6 (x_4-x_6) \nn \\
&& +\, \f{(u^-_1u^+_3)}{(u^+_1u^+_3)} \delta^{I_2 L_1} f^{ K_2  N_3
L_2} \f{1}{\square^2_4} \delta^{6} (x_4-x_6) \f{1}{\square_4}
\delta^6 (x_4-x_5)\bigg)
 \nn \\
&& \times \f{1}{\square_1 }\delta^6(x_1-x_4)
\f{1}{\square_1}\delta^6(x_1-x_5)\f{1}{\square_1}\delta^6(x_1-x_6)
\,. \label{GammaVII-3}
\end{eqnarray}
Then we replace the space-time variables $ x_5$ by $x_6 $ and the
harmonic ones $u_2$ by $u_3$ in the second term inside the bracket.
As a result, the  expression considered can be rewritten as
  \begin{eqnarray}
\Gamma^{FQQ}_{\rm VII,\, div} &=& 16 {\rm f}^2 f^{I_1J_1K_1}
f^{I_1I_2K_2}
f^{L_1J_1N_1} f^{K_2 K_1 N_2} f^{I_2 N_3 L_1} \int d^{14}z_1 d^{6}x_4 d^{6}x_5 d^{6}x_6 d u_1 du_2 du_3 \nn\\
&&\times \f{(u^-_1u^+_2)}{(u^+_1u^+_2)(u^+_2u^+_3)} \,
 (Q^{+A})^{N_1}(x_5,\theta_1,u_2) (Q^{+}_{A})^{N_2}(x_6,\theta_1,u_3) (F^{++})^{N_3}(x_4,\theta_1,u_1)  \nn\\
&&  \times \f{1}{\square^2_4} \delta^{6} (x_4-x_5) \f{1}{\square_4}
\delta^6 (x_4-x_6)
 \f{1}{\square_1 }\delta^6(x_1-x_4) \f{1}{\square_1}\delta^6(x_1-x_5)\f{1}{\square_1}\delta^6(x_1-x_6)
\,. \label{GammaVII-4}
\end{eqnarray}
Calculating the momentum integral employing \eqref{momint} and
taking the trace over the gauge algebra indices, we obtain
 \begin{equation}\label{GammaVII-5}
 \Gamma^{\rm FQQ}_{\rm V II,\, div} = -\f{4i{\rm f}^2 (C_2)^2}{(4\pi)^6 \varepsilon^2} \tr
 \int d^{14}z  \f{d u_1 du_2 du_3(u_1^-u_2^+)}{(u_1^+u_2^+)(u_2^+u_3^+)} F^{++}_{1,\tau} [Q^{+A}_{2,\tau}, Q^+_{3A,\tau}]\,.
\end{equation}

\paragraph{The final result for divergent contributions.}
Summing up the divergent contributions \eqref{GammaI},
\eqref{GammaV-2} and \eqref{GammaVII-5} and singling out the
structure we are interested in, we obtain
 \bea\label{NonLocalResult}
 \Gamma^{\rm FQQ}_{\rm I, \,div} + \Gamma^{\rm FQQ}_{\rm V,\, div} + \Gamma^{\rm FQQ}_{\rm VII,\, div}
&=& -\f{i{\rm f}^2(C_2)^2}{8(2\pi)^6\,\varepsilon^2}\tr \int d^{14}z
\, \f{du_1 du_2 du_3}{(u^+_1 u^+_2)(u^+_3 u^+_1)} F^{++}_{1, \tau}
[Q^{+A}_{2,\tau}, Q^{+}_{3 A,\tau}] \nn\\
&& +\f{2i{\rm f}^2 (C_2)^2}{(4\pi)^6 \varepsilon^2} \tr
 \int d^{14}z \f{d u_1 du_2}{(u_1^+u_2^+)^2} F^{++}_{1,\tau} [Q^{+A}_{2,\tau}, Q^+_{2A,\tau}] \nn\\
&& - \f{4i{\rm f}^2 (C_2)^2}{(4\pi)^6 \varepsilon^2} \tr
 \int d^{14}z  \f{d u_1 du_2 du_3(u_1^-u_2^+)}{(u_1^+u_2^+)(u_2^+u_3^+)} F^{++}_{1,\tau} [Q^{+A}_{2,\tau}, Q^+_{3A,\tau}]
 \eea
This expression agrees with the power counting analysis performed
above. We see that, as expected, the divergent contributions to the
effective action are non-local in harmonics, but remain local with
respect to the space-time and Grassmann coordinates.

\paragraph{On-shell condition.}

In Appendix \ref{appendix} the expression (\ref{NonLocalResult}) is
rewritten in a different form that is local in the harmonic
variables. However, the result involves the superfield $Q^{-A}$
defined by the equation $\nabla^{++} Q^{-}_{A} = Q^{+}_A$.
Nevertheless, all terms containing $Q^{-}_A$ vanish on shell. Then
it becomes possible to determine the coefficient $c_2$ in
\eqref{Gdiv2}, and also to fix the last coefficient $c_3$ in
\eqref{Gdiv2} using eq. (\ref{c3equation}):
 \bea
 c_2= - \frac{{\rm f^2} (C_2)^2}{16(2\pi)^6\varepsilon^2}\,, \qquad
 c_3=\frac{{\rm f^2} (C_2)^2}{16(2\pi)^6\varepsilon^2}\,, \qquad
 \varepsilon \to 0
 \label{coefic2}\,.
\eea The final expression for the divergent part of the two-loop
effective action in the model \eq{S0} reads \bea
 \Gamma^{(2)}_{\rm div} &=& \frac{{\rm f^2} (C_2)^2}{8(2\pi)^6\varepsilon^2} \tr \int d \zeta^{(-4)}\Big(
  F^{++} \sB F^{++} -\frac{i}{2} \, F^{++} \sB [Q^{+A},Q^+_A] \nn \\
  && + \frac12 \, [Q^{+A},Q^+_A] \sB [Q^{+B},Q^+_B] \Big)
 + \text{terms proportional to e.o.m. for $Q^+$}.
 \label{Gdivfin}
 \eea
The expression \eqref{Gdivfin} vanishes for the background
superfields satisfying the classical equations of motion \eqref{eqm}
and so generalizes the result obtained earlier in ref.
\cite{Buchbinder:2021unt}.


\section{Summary}\label{Section_Summary}

In the present paper we have developed a manifestly covariant and
$\cN=(1,0)$ supersymmetric method for studying the effective action
of the six-dimensional $\cN=(1,1)$ SYM theory formulated in
$\cN=(1,0)$ harmonic superspace. In such an approach the classical
action (\ref{S0}) is that of ${\cal N}=(1,0)$ SYM theory minimally
interacting with the hypermultiplet in the adjoint representation of
the gauge group. The corresponding classical action is invariant
under ${\cal N}=(1,0)$ supersymmetry and also possesses a hidden
on-shell $\cN=(0,1)$ supersymmetry. Hence, the action \eq{S0} really
describes the six-dimensional $\cN=(1,1)$ SYM theory.

In the previous papers \cite{Buchbinder:2016url,Buchbinder:2016wng}
we have shown that in the minimal gauge (a supersymmetric version of
the Feynman gauge) this theory is finite off shell in the one-loop
approximation. The two-loop $\frac{1}{{\varepsilon}^2}$ divergent
contribution in the gauge-multiplet sector has been calculated in
ref. \cite{Buchbinder:2021unt}. In the present paper we have
analyzed the general structure of the two-loop divergences in the
theory under consideration and shown that, generically, the
corresponding counterterms, being space-time local, can be non-local
in harmonics. Extending the result of ref.
\cite{Buchbinder:2021unt}, we have performed the explicit
calculation of the leading two-loop divergences in the mixed gauge
multiplet-hypermultiplet sector. Note that the calculation of the
divergences containing the background hypermultiplet has required
considerable efforts with the use of a large arsenal of various
harmonic identities.

Throughout the paper we performed the calculation within the
background field method in the harmonic superspace for arbitrary
background superfields $V^{++}$ and $Q^{+A}$. Due to the presence of
the non-trivial hypermultiplet background the kinetic term for
quantum superfields contains the mixed term that is removed by the
non-local shift \eqref{replac} of the quantum hypermultiplet. This
leads to the appearance of new cubic vertices \eqref{vert5} and
\eqref{vert6} and to the new two-loop diagrams presented in Fig.
\ref{Fig3}, some of which contribute to the divergent part of the
effective action. Also, the two-loop divergences come from the
supergraph $\Gamma_{\rm I}$ considered earlier in
\cite{Buchbinder:2021unt} under the choice $Q^+=0$. In this paper we
carried out a direct calculation of the two-loop
$\frac{1}{{\varepsilon}^2}$ divergences and showed that the final
result, with contributions from both the gauge multiplet and
hypermultiplet, exactly coincides with the one obtained in ref.
\cite{Buchbinder:2021unt}, where the dependence on the
hypermultiplet was restored through an indirect analysis.

It is worth noting that, in addition to the leading
$\frac{1}{{\varepsilon}^2}$ divergences, one can also expect the
two-loop subleading $\frac{1}{\varepsilon}$ divergences. If the
background hypermultiplet is switched off, the expressions  for them
are proportional to $F^{++}$ and so naturally vanishes on shell due
to the equations of motion for the vector multiplet. However, the
calculation of the $\frac{1}{\varepsilon}$ divergences is much more
technically involved as compared to the $\frac{1}{{\varepsilon}^2}$
terms (especially in the presence of the background hypermultiplet)
and the further work is needed to study the structure of the
background-dependent harmonic supergraphs. It would be very
interesting (and instructive from the point of view of the harmonic
supergraph technique) to derive the total set of the
$\frac{1}{\varepsilon}$ divergences by a direct calculation. We are
going to study all the relevant aspects in the forthcoming works.
However, we wish here to emphasize that all calculations can be
carried out using the same techniques. For illustration, in Appendix
B we show how the harmonic technique works for calculating the
two-loop $\frac{1}{\epsilon}$ divergences of the supergraph
presented in Fig. \ref{Fig3}.

As we saw, the one- and two-loop divergences in the theory under
consideration are proportional to classical equations of motion.
This implies the one- and two-loop on-shell finiteness of the theory
under consideration. One can expect that, starting from the
three-loop approximation, the structure of divergences will change
and the results will not be proportional to the standard classical
equations of motion. At present, the off-shell structure of the
three- and higher-loop divergences of the effective action is
unknown\footnote{The existence of the proper dimension single-trace
operators with the manifest off-shell ${\cal N}=(1,0)$ supersymmetry
and hidden on-shell ${\cal N}=(0,1)$ supersymmetry was shown in
\cite{Bossard:2015dva} (see also \cite{SerIvan}). These operators
are non-vanishing on shell and so can carry actual divergences.}. We
plan to address this problem in the future papers.

\section*{Acknowledgements}

Work of I.L.B., E.A.I. and K.V.S. was supported by Russian
Scientific Foundation, project No 21-12-00129. Work of B.S.M was
supported in part by the Ministry of Education of the Russian
Federation, project No QZOY-2023-0003.

\bigskip

\appendix

\section{Harmonic locality} \label{appendix}

The contributions \eqref{dim3_1}, \eqref{dim5}, \eqref{dim6} and
\eqref{dim7} to the two-loop effective action are non-local in the
harmonic variables. This does not contradict the non-renormalization
theorem which asserts only the locality of the divergent part of
effective action in the Grassmann and space-time coordinates.
Moreover, all the contributions just mentioned can be related to
local expressions such as \eqref{dim2}, \eqref{Gam2}. Here we are
going to demonstrate this.

Let us begin our study with
 \bea
 {\cal G}^{(2)}_{1} = \tr\int d^{14}z \f{du_1du_2}{(u_1^+u_2^+)^2}
 F^{++}_{1,\tau} F^{++}_{2,\tau}\,,
 \label{dim3appendix}
 \eea
where $F^{++}_{1, \tau}=F^{++}_{\tau}(z,u_1)$, $[F^{++}]=[m^2]$ and
$\tau$-representation was introduced in Eq. \eqref{tau}. Using the
identity
 \bea
 \f{1}{(u^+_1u^+_2)^2} = D^{++}_1  \f{(u_1^-u_2^+)}{(u_1^+ u_2^+)^3}
 + \f12 (D^{--}_1)^2 \d^{(2,-2)}(u_1,u_2)
 \label{id3}
 \eea
this expression can be rewritten as
 \bea
 {\cal G}^{(2)}_{1} = \tr\int d^{14}z du_1du_2\Big( D^{++}_1 \f{(u_1^-u_2^+)}{(u_1^+ u_2^+)^3}
 + \f12 (D^{--}_1)^2 \d^{(2,-2)}(u_1,u_2) \Big)
 F^{++}_{1,\tau} F^{++}_{2,\tau}\,.
  \eea
The first term in the bracket is a total $D^{++}$ harmonic
derivative due to the property $\nabla^{++} F^{++} =0$. In the
second term we integrate by parts with respect to the derivative
$D_1^{--}$ and then use the harmonic delta function to integrate
over $u_2$. We obtain
 \bea
 {\cal G}^{(2)}_{1} =  \f12  \tr\int d^{14}z\, du \,
 F_{\tau}^{++}(D^{--})^2 F_{\tau}^{++}\, =  \tr\int d\zeta^{(-4)}\,
 F^{++} \sB F^{++}\,.
  \eea
We see that this expression is invariant under the gauge
transformation \eqref{gtr} (the initial expression which is
non-local in the harmonic variables is evidently gauge invariant
too).

Now let us consider the expression \eqref{dim6},
\begin{eqnarray}
{\cal G}^{(2)}_2 &=& \tr \int d^{14} z \f{du_1 du_2}{(u_1^+
u_2^+)^2}
 F^{++}_{1,\tau}[Q^{+A}_{2,\tau}, Q^+_{2A,\tau}] \nn \\
&=& \tr \int d^{14} z du_1 du_2 \Big( D^{++}_1
\f{(u_1^-u_2^+)}{(u_1^+ u_2^+)^3} + \f12 (D^{--}_1)^2
\d^{(2,-2)}(u_1,u_2) \Big)
 F^{++}_{1,\tau}[Q_{2,\tau}^{+A},Q^+_{2 A,\tau}]\,.
\end{eqnarray}
The first term in the bracket is the total harmonic derivative
$D^{++}$ due to the property $\nabla^{++} F^{++}=0$. In the second
term we integrate twice by parts  with respect to the harmonic
derivative $D^{--}$,
\begin{eqnarray}
{\cal G}^{(2)}_2 &=& \f12  \tr \int d^{14}z [Q^{+A},Q^+_{A}]
(\nabla^{--})^2 F^{++} = \tr \int d\zeta^{(-4)} [Q^{+A},Q^+_{A}] \sB
F^{++} \,.
\end{eqnarray}

The expression \eqref{dim7} can equivalently be presented in the
form
\begin{equation}\label{second_expression}
{\cal G}^{(2)}_3 = \tr \int d^{14} z \f{du_1 du_2 du_3}{(u_1^+
u_2^+) (u_1^+u_3^+)} F^{++}_{1,\tau}[Q_{2,\tau}^{+A},Q^+_{3 A,\tau}]
= \tr \int d^{14} z du\, F^{++}[Q^{-A},Q^-_{A}]\,,
\end{equation}
where the superfield $Q^{-}_A$ is defined by the equation
$\nabla^{++} Q^-_{A} = Q^+_A$. Its solution in the
$\tau$-representation is written as
 \bea\label{QMinus}
Q^{-}_{A,\tau}(z,u_1) = \int du_2
\f{Q^{+}_{A,\tau}(z,u_2)}{(u^+_1u^+_2)}\,.
 \eea
Then, using the identity
 \bea
 Q^-_A= \nabla^{--}Q^+_A - \nabla^{++}\nabla^{--} Q^-_A\,,
 \eea
we rewrite the expression (\ref{second_expression}) in the form
 \bea
&& {\cal G}^{(2)}_3 = \tr \int d^{14}z\, du\, \Big(
F^{++}[\nabla^{--}Q^{+A},Q^-_{A}] -
F^{++}[\nabla^{++}\nabla^{--}Q^{-A},Q^-_{A}]\Big) \nn \\
&& = -\tr \int d^{14}z\, du \, [Q^{+A},Q^-_{A}] \nabla^{--}F^{++} \nn \\
&& = -\tr \int d^{14}z\, du \, \Big([Q^{+A},\nabla^{--}Q^+_{A}]
-[Q^{+A},\nabla^{++}\nabla^{--}Q^-_{A}]\Big)\nabla^{--}F^{++} \nn\\
&& = -\tr \int d^{14}z\, du \, \Big(\tfrac12
\nabla^{--}[Q^{+A},Q^+_{A}]
-[Q^{+A},\nabla^{++}\nabla^{--}Q^-_{A}]\Big)\nabla^{--}F^{++} \nn \\
&& = \frac12 \tr \int d^{14}z\, du \, F^{++}
(\nabla^{--})^2[Q^{+A},Q^+_{A}] + \tr \int d^{14}z\, du
\,[Q^{+A},\nabla^{++}\nabla^{--}Q^-_{A}]
\nabla^{--}F^{++} \nn \\
&& = \tr \int d \zeta^{(-4)}\, F^{++} \sB [Q^{+A},Q^+_{A}] + \tr
\int d^{14}z\, du \,[Q^{+A},\nabla^{++}\nabla^{--}Q^-_{A}]
\nabla^{--}F^{++}\,.
 \label{dim8}
 \eea
In the first line we integrated by parts with respect to  the
harmonic derivatives $\nabla^{--}$ and $\nabla^{++}$ in the first
and second terms, respectively. The first term in the last line of
eq. \eq{dim8} coincides with \eqref{Gam2}. The second term vanishes
if the background hypermultiplet satisfies the classical equation of
motion \eqref{eqm} (this is because $\nabla^{++} \nabla^{--} Q^-_A =
0$ on shell, see ref. \cite{Buchbinder:2018lbd} for details).

Let us consider the last term  \eqref{dim7}
 \bea
 {\cal G}_{4}^{(2)} &=& \tr
 \int d^{14}z  \f{d u_1 du_2 du_3\,(u_1^-u_2^+)}{(u_1^+u_2^+)(u_2^+u_3^+)}
 F^{++}_{1,\tau} [Q^{+A}_{2,\tau}, Q^+_{3A,\tau}]\,.
 \label{dim9}
  \eea
Using eq. (\ref{QMinus}) and the identity
 \bea
\f{(u_1^+u_2^+)}{(u_1^+u_2^+)} = \f12 D^{++}_1
\f{(u_1^+u_2^+)^2}{(u_1^+u_2^+)^2} - \f{1}{2}D^{--}_1
\delta^{(0,0)}(u_1,u_2)\,,
 \eea
we rewrite this expression, modulo a complete harmonic derivative,
as
 \bea
 {\cal G}_{4}^{(2)} &=& -\f12 \tr \int d^{14}z\, d u_1 du_2 \,D^{--}_1 \delta^{(0,0)}(u_1,u_2)
 F^{++}_{1,\tau}[Q^{+A}_{2,\tau}, Q^-_{2A,\tau}] \nn \\
 &=& -\f14 \tr \int d^{14}z\, d u_1 du_2 \,D^{--}_1 \delta^{(0,0)}(u_1,u_2)
 F^{++}_{1,\tau} D^{--}_2 [Q^{+A}_{2,\tau}, Q^+_{2A,\tau}] \nn \\
&& +  \f12 \tr \int d^{14}z\, d u_1 du_2 \,D^{--}_1
\delta^{(0,0)}(u_1,u_2)
 F^{++}_{1,\tau} D^{--}_2 [Q^{+A}_{2,\tau}, D^{++}_2 D^{--}_2 Q^-_{2A,\tau}] \nn \\
 &=& \f14 \tr \int d^{14}z\, d u\,
\nabla^{--} F^{++} \nabla^{--}  [Q^{+A}, Q^+_{A}] \nn \\
&& -  \f12 \tr \int d^{14}z d u \nabla^{--}
 F^{++}\nabla^{--}  [Q^{+A}, \nabla^{++}\nabla^{--} Q^-_{A}]\nn \\
 && = - \f12 \tr \int d\zeta^{(-4)}\,
F^{++} \sB [Q^{+A}, Q^+_{A}] \nn \\
&&  -  \f12 \tr \int d^{14}z\, d u \nabla^{--}
 F^{++}\nabla^{--}  [Q^{+A}, \nabla^{++}\nabla^{--} Q^-_{A}]\,.
 \label{dim9}
 \eea
Like in the previous case, the second term in the last line of
eq.\eqref{dim9} vanishes on shell because of the equality
$\nabla^{++} \nabla^{--} Q^-_A = 0$.

Acting similarly, one can show that the non-local contribution
\eqref{dim10} coincides with \eqref{Gam3} up to terms proportional
to the background hypermultiplet equation of motion.

Finally, we note that all non-local in the harmonics expressions
\eqref{dim3_1}, \eqref{dim5}, \eqref{dim6} and \eqref{dim7} are
gauge invariant. In the $\lambda$-representation, this directly
follows from the above analysis.

\section{Calculation of $\frac{1}{\epsilon}$ divergences: an example}\label{appendix2}
In this Appendix we illustrate how the two-loop
$\frac{1}{\varepsilon}$ divergences can in principle be calculated
in the framework of the harmonic supergraph techniques. As an
example, the consideration will be carried out for the supergraph
presented in Fig. \ref{Fig3}.

It is known that the theory under consideration does not have
one-loop divergences off shell at least in the Fermi-Feynman gauge
\cite{Buchbinder:2017xjb}, hence no sub-divergences in the
two-loop supergraphs can appear. For this reason, the divergent contributions
proportional to $\frac{1}{\varepsilon}$ can come out  only as the cross terms, with
the divergent contributions multiplied by the finite ones. However, all
such contributions also vanish on shell. To demonstrate this, we  consider, for simplicity,
the trivial hypermultiplet background $Q^+=0$. In
this case  all additional diagrams with ''non-local'' vertices in
Fig. \ref{Fig3}, including the diagram $\Gamma_{\rm V}$ which
contains the divergent contribution, immediately vanish.
Consequently, all possible divergent contributions can come only from the
diagram $\Gamma_{\rm I}$.

Let us denote $g^{(1,1)}(z; u_1,u_2)$ the finite contribution of
the vector multiplet Green function divided by $(u_1^+u_2^+)$
\eq{Gg4} in the $z_2 \to z_1$ limit,
 \bea
\f{G^{(2,2)}(z_1,u_1|z_2,u_2)}{(u_1^+u_2^+)}\Big|_{z_2\to z_1} =
\f{a}{\varepsilon}\, F^{++}_1 \,(u_1^-u_2^+)^2
\delta^{(1,-1)}(u_1,u_2) + g^{(1,1)}(z; u_1,u_2|V^{++})\,, \quad
\varepsilon\to 0\,. \label{Gg5}
 \eea
Here, $a=-\f{4i}{(4\pi)^3}$. The function $g^{(1,1)}(z;u_1,u_2|V^{++})$
is an analytic superfield depending on
the background vector multiplet $V^{++}$. By definition, this function
is a gauge invariant polynomial of
gauge superfield strength, such that it is local in the $z$ space and can be non-local
with respect to harmonic
variables. It contains all the information about the structure of the
finite part of the gauge superfield Green function $G^{(2,2)}(z_1,
u_1;z_2, u_2)$ in the $z_2\to z_1$ coincident limit. The explicit form
of this function does not matter and will not be discussed in what follows.

As we discussed earlier,  the divergent contribution of the diagram
$\Gamma_{\rm I}$ arises from the expression (we omitted the gauge group
indices and relevant coefficients)
 \bea
\Gamma^{(2)}_{\rm div}[V^{++}] &=& {\rm f}^2\int d^{14}z
\prod_{a=1}^{4} u_a \frac{G^{(2,2)}(z,u_1 ; z, u_2) \, G^{(2,2)} (z,
u_3; z, u_4) } {(u^+_1u^+_2)(u^+_2u^+_3)(u^+_3 u^+_4) (u^+_4
u^+_1)}\bigg|_{\rm div}\,.
 \eea
Substituting \eq{Gg5} in the last expression and integrating with the
harmonic delta-functions, we obtain
 \bea
\Gamma^{(2)}_{\rm div}[V^{++}] &=&-\f{a^2{\rm
f^2}}{\varepsilon^2}\tr \int d\zeta^{(-4)} \, F^{++} \sB F^{++} \nn
\\
&& +\f{2a {\rm f^2}}{\varepsilon}\tr \int d^{14}z \f{du_1 du_2
du_3}{(u_1^+u_2^+)(u_1^+u_3^+)}F^{++}_{1, \tau}
g^{(1,1)}(z;u_2,u_3|V^{++})\,. \label{GammaI2}
 \eea
The first term in \eq{GammaI2} was obtained earlier in
\cite{Buchbinder:2021unt} by explicit calculation of the effective
action, and it corresponds to the highest-order pole in $\varepsilon$. The
second term corresponds to the $\f{1}{\varepsilon}$ divergent
contribution. We see that such a contribution also vanishes when the
background gauge superfield satisfies the classical equation of
motion, $F^{++}=0$, in agreement  with the hidden $\cN=(0,1)$
supersymmetry.

Thus we see that the harmonic supergraph technique successfully works
for calculating both $\frac{1}{{\varepsilon}^2}$ and
$\frac{1}{{\varepsilon}}$ divergences.

\end{document}